\documentclass{aa}
\usepackage[varg]{txfonts}
\usepackage{natbib}
\usepackage{graphicx}
\usepackage{float}
\usepackage{lscape}
\usepackage{caption}
\usepackage{subcaption} 
\usepackage{sidecap}
\usepackage{xcolor}
\usepackage{comment}
\titlerunning{}
\usepackage{tikz}
\usetikzlibrary{positioning}
\begin{document} 

   \title{ Three New Galactic Globular Cluster Candidates: FSR1700, Teutsch67, and CWNU4193}
   \author{}
   \author{Saroon S
          \inst{1}\fnmsep\thanks{saroonsasi19@gmail.com}
           \and
           Bruno Dias\inst{1}
           \and
           Dante Minniti\inst{1,2,3} 
           \and
           M. C. Parisi\inst{4,5}
           \and 
           Matías Gómez \inst{1}
            \and
           Javier Alonso-García \inst{6,7}
           }
          
   \institute{
$^{1}$ Instituto de Astrofísica, Departamento de Ciencias Físicas, Facultad de Ciencias Exactas, Universidad Andres Bello, Fernández Concha 700, Las Condes, Santiago, Chile.  \\ 
   $^{2}$ Vatican Observatory, V00120 Vatican City State, Italy.\\
   $^{3}$Departamento de Fisica, Universidade Federal de Santa Catarina, Trinidade 88040-900, Florianopolis, Brazil.\\
   $^{4}$Observatorio Astronómico, Universidad Nacional de Córdoba, Laprida 854, X5000BGR, Córdoba, Argentina.\\
   $^{5}$Instituto de Astronomía Teórica y Experimental (CONICET-UNC), Laprida 854, X5000BGR, Córdoba, Argentina.\\
   $^{6}$ Centro de Astronom\'{i}a (CITEVA), Universidad de Antofagasta,
  Av. Angamos 601, Antofagasta, Chile  \\
  $^{7}$Millennium Institute of Astrophysics, Nuncio Monse\~nor Sotero Sanz 100,
  Of. 104, Providencia, Santiago, Chile
            } 
   \date{Received---- ; accepted----}
  \abstract
   { 
The VISTA Variables in the Via Láctea Extended Survey (VVVX) enables exploration of previously uncharted territories within the inner Milky Way (MW), particularly those obscured by stellar crowding and intense extinction. 
Our objective is to identify and investigate new star clusters to elucidate their intrinsic characteristics. Specifically, we are focused on uncovering new candidate Globular Clusters (GCs) situated at low Galactic latitudes, with the ultimate goal of completing the census of the MW GC system. Leveraging a combination of Near-InfraRed (NIR) data from the VVVX survey and Two Micron All Sky Survey (2MASS), along with optical photometry and precise proper motions (PMs) from the Gaia Data Release 3 (DR3), we are conducting a systematic characterisation of new GCs. 
As a result, we report the discovery and characterisation of four new Galactic clusters named FSR~1700, FSR~1415, CWNU~4193, and Teutsch~67, all located within the MW disk.

We estimate a wide range of reddening, with values ranging from 0.44 to 0.73 mag for $E(J-Ks)$. The heliocentric distances span from 10.3 to 13.2 kpc. Additionally, we determine their metallicities and ages, finding a range of -0.85 to -0.75 dex for [Fe/H] and ages approximately close to 11 Gyr, respectively. FSR~1415 is an exception, it is an old open cluster with age = 3 Gyr and [Fe/H] = -0.10. Furthermore, we fitted the radial density profiles to derive their structural parameters like tidal radius, core radius, and concentration parameters. In conclusion, based on their positions, kinematics, metallicities, and ages, and comparing our findings with existing literature, we categorise FSR~1700, Teutsch~67 and CWNU~4193 as genuine GC candidates, while FSR~1415 is an old open cluster exhibiting characteristics of a post core-collapse cluster.

   }
    \keywords{Galaxy:disk–Galaxy:stellar content–globular clusters:general–infrared:stars–surveys }
    
\maketitle

\begin{figure*}[t]
    \centering
    \begin{subfigure}[b]{0.45\textwidth}
        \centering
        \includegraphics[width=\textwidth]{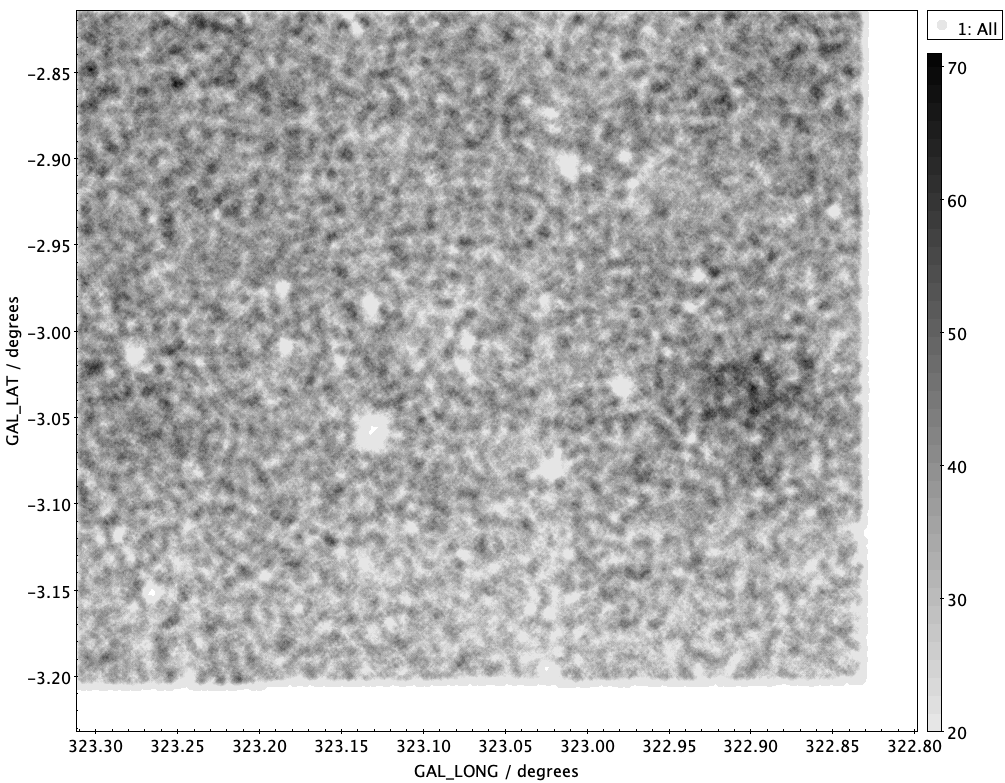}
        \caption{FSR1700}
        \label{fig:sub1}
    \end{subfigure}
    \begin{subfigure}[b]{0.45\textwidth}
        \centering
        \includegraphics[width=\textwidth]{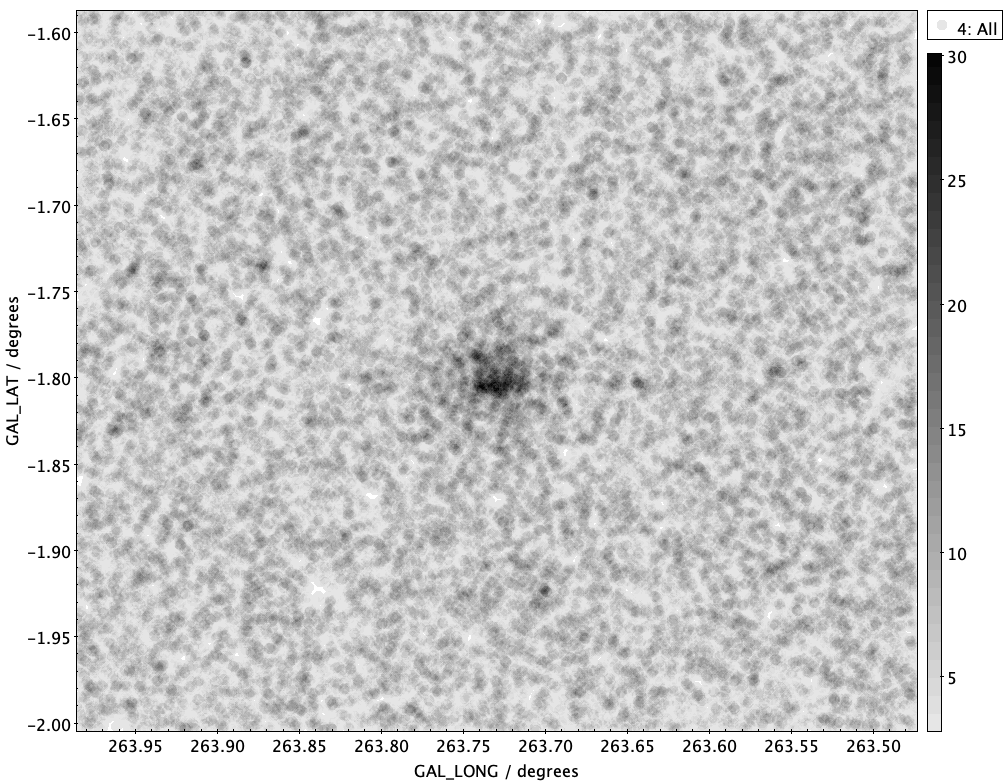}
        \caption{FSR1415}
        \label{fig:sub2}
    \end{subfigure}
    \begin{subfigure}[b]{0.45\textwidth}
        \centering
        \includegraphics[width=\textwidth]{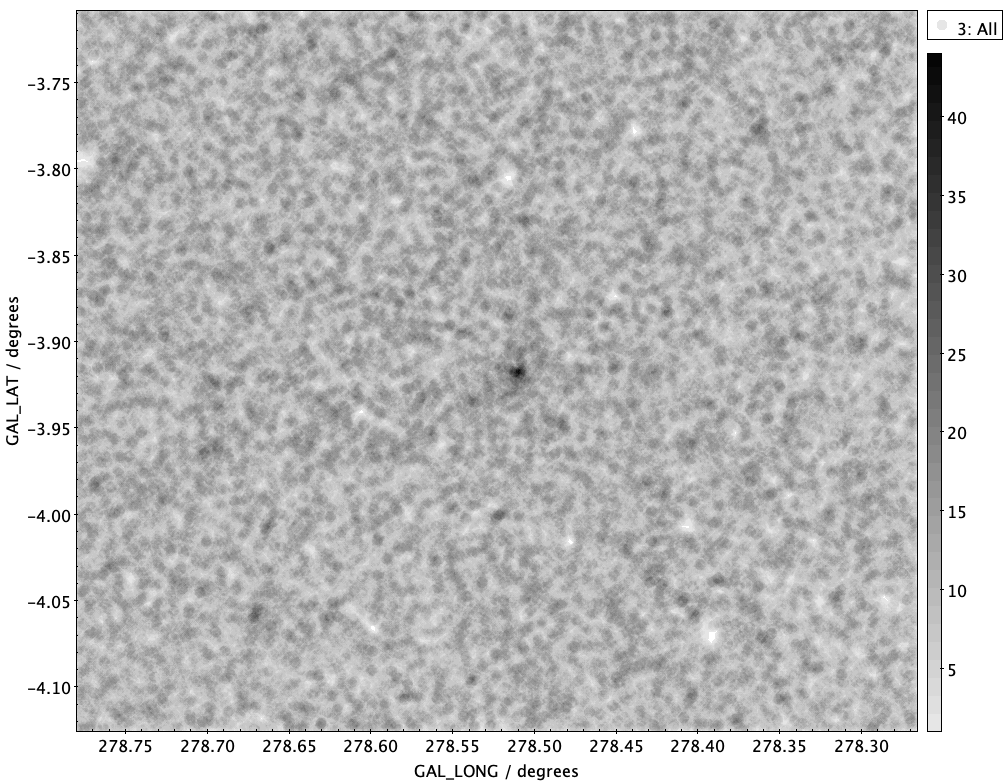}
        \caption{Teutsch67}
        \label{fig:sub3}
    \end{subfigure}
    \begin{subfigure}[b]{0.45\textwidth}
        \centering
        \includegraphics[width=\textwidth]{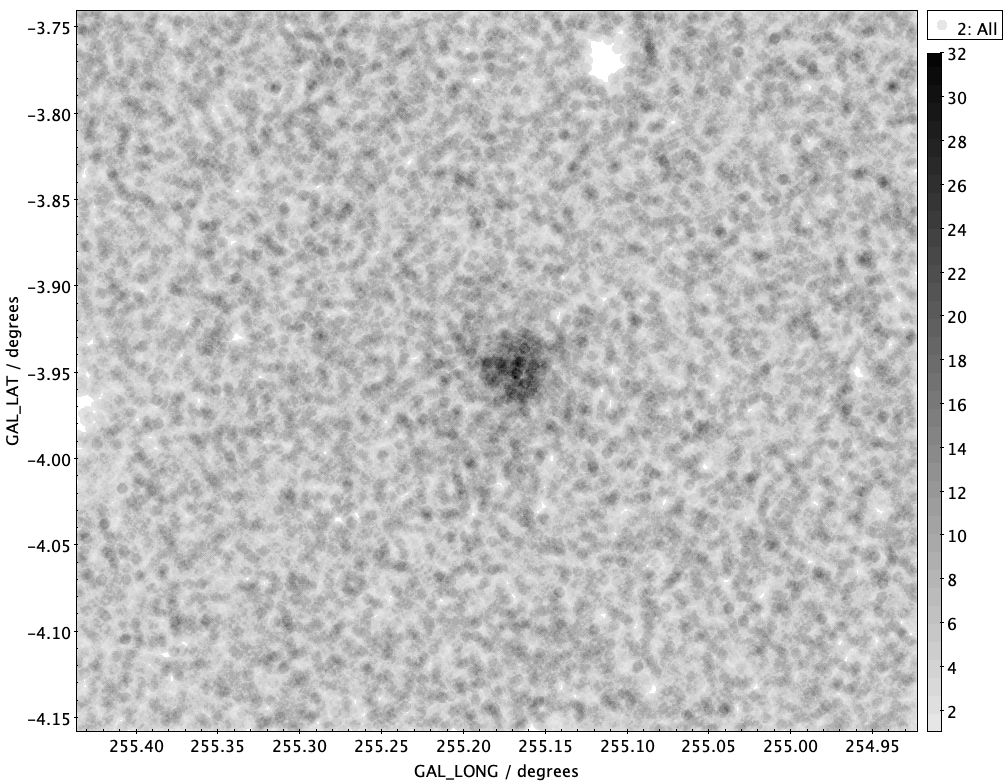}
        \caption{CWNU4193}
        \label{fig:sub4}
    \end{subfigure}
    \caption{Density map of the observed clusters using the VVVX data. The bar shows the grey scale corresponding to the number density of stars in a linear scale.}
    \label{ImgDmap}
\end{figure*}

\section{Introduction}

The Milky Way environment serves as a distinctive testing ground for evaluating the predictions of cosmological models and theories of galaxy formation. Additionally, our vantage point within our own Galaxy affords us an intimate examination of old
star clusters, bearing implications for extra-galactic star cluster studies. These insights become particularly significant with the advent of next-generation facilities such as NASA's James Webb Space Telescope (JWST) and the upcoming ESO's European Extremely Large Telescope (ELT). 

Over the past decade, a plethora of new Globular Cluster (GC) candidates has emerged throughout the Milky Way (MW), particularly in the bulge regions where interstellar dust obscures baryonic matter extensively across the sky (e.g., \citealp{2012Gonzalez}). The challenging conditions, marked by high stellar crowding, have prompted intensified exploration. Initiatives like the VISTA Variables in the Via Lactea (VVV; \citealp{2010MinnitiVVV,2012Saito}) and its extension, VVVX (\citealp{VVVXDM2018}), have significantly augmented the catalog of star cluster candidates in the MW. These investigations employ either visual inspection or photometric analysis to unveil previously unknown star clusters, showcasing remarkable success in these endeavours (e.g., \citealp{2019+Bica,2021+DM}).

\cite{2017DM} utilised density maps constructed solely from red giants, identifying apparent overdensities indicative of GC candidates. These overdensities were visually inspected, considering their size relative to known Galactic GCs ($\sim 2' - 5'$), and compared the Color Magnitude Diagrams (CMDs) of potential candidates with those of well characterised Galactic GCs and their respective background fields. However, not all overdensities signify true clusters; some may merely constitute groups of stars or statistical fluctuations in the projected stellar density on the sky plane (\citealp{2019Gran,2019Palma,2021DMinniti}). 
Hence, one of the most dependable methods for confirming or refuting the existence of a cluster is kinematic analysis using the high precision of 
VVV (\citealp{smith2018virac})
and Gaia (\citealp{GDR32023}) Proper Motions (PMs; e.g.,\citealp{2020Garro,2021casmir,2021+DM}).
In this work we report the discovery and characterisation of four old star
clusters: FSR~1700, FSR~1415, CWNU~4193, and Teutsch~67 which are embedded in the Galactic plane, away from the bulge.

The paper is organised as follows: In Section 2, we provide a brief overview of the datasets employed in this study. Section 3 outlines the decontamination procedure, delineating the process of extracting the cluster from the field population. The methods used to estimate the astrophysical and structural parameters are elucidated in Section 4. In Section 5, we present the results and conduct a comparative analysis with existing literature for each cluster. Finally, a summary of our findings and conclusions are presented in Section 6.
\begin{figure*}[t]
    \centering
    \begin{subfigure}[b]{0.45\textwidth}
         \includegraphics[width=\textwidth]{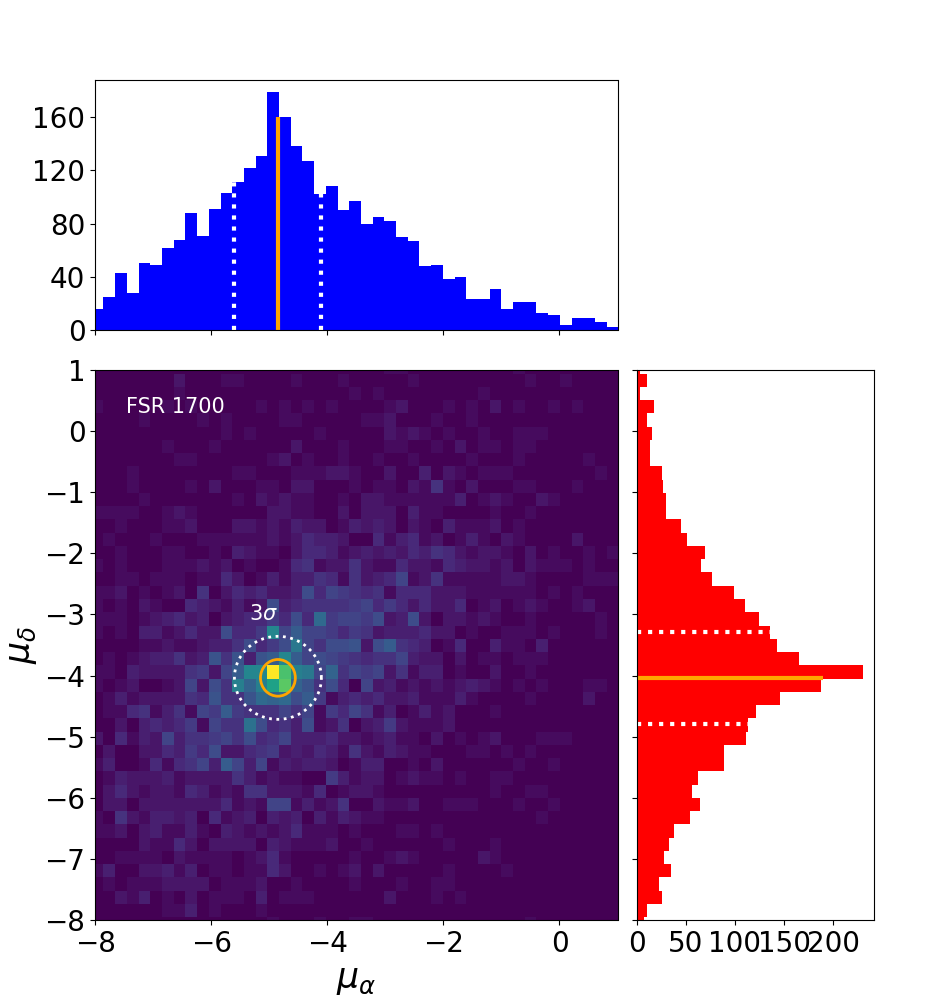}
     \end{subfigure}
      \begin{subfigure}[b]{0.45\textwidth}
         \includegraphics[width=\textwidth]{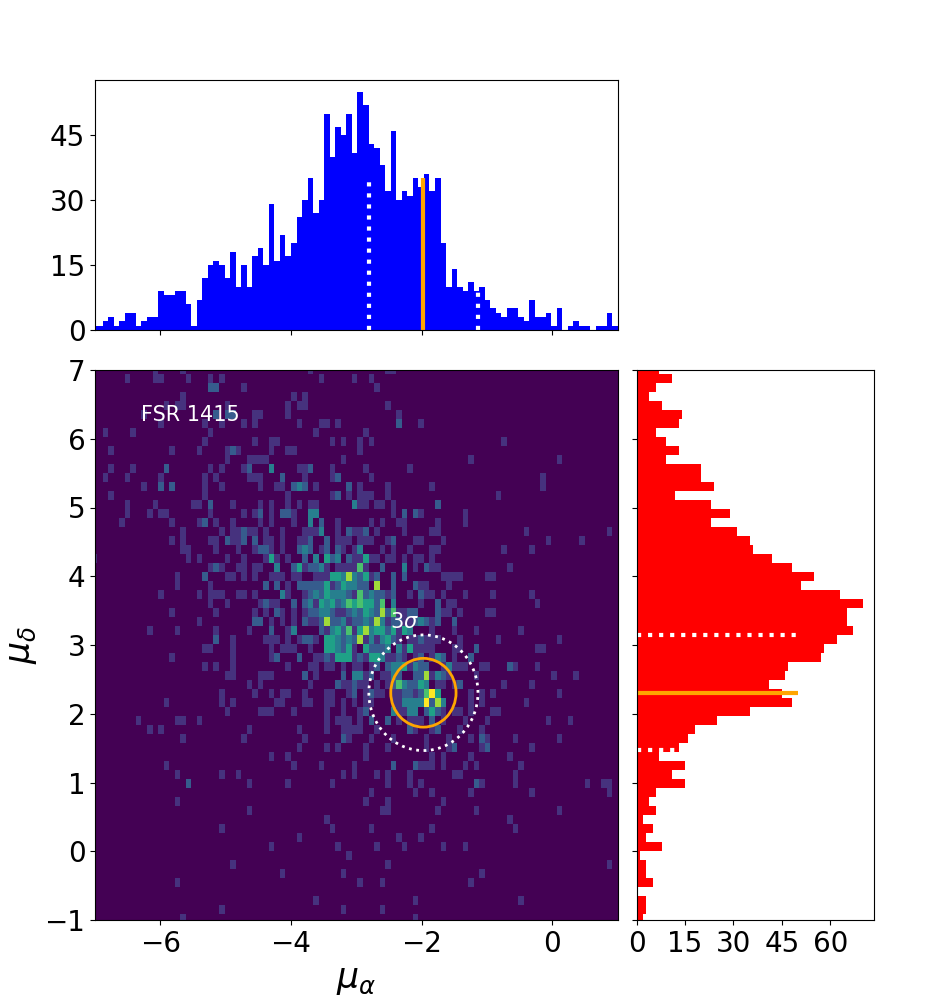}
     \end{subfigure}
     \begin{subfigure}[b]{0.45\textwidth}
         \includegraphics[width=\textwidth]{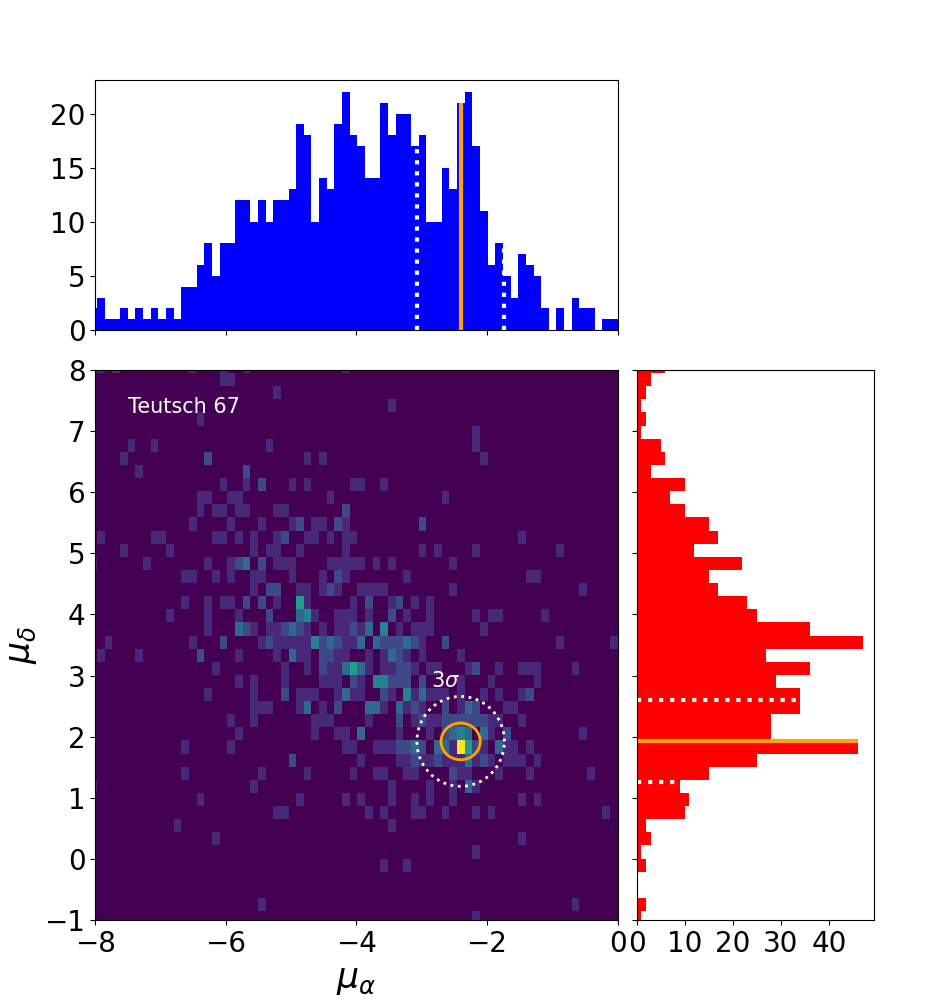}
     \end{subfigure}
      \begin{subfigure}[b]{0.45\textwidth}
         \includegraphics[width=\textwidth]{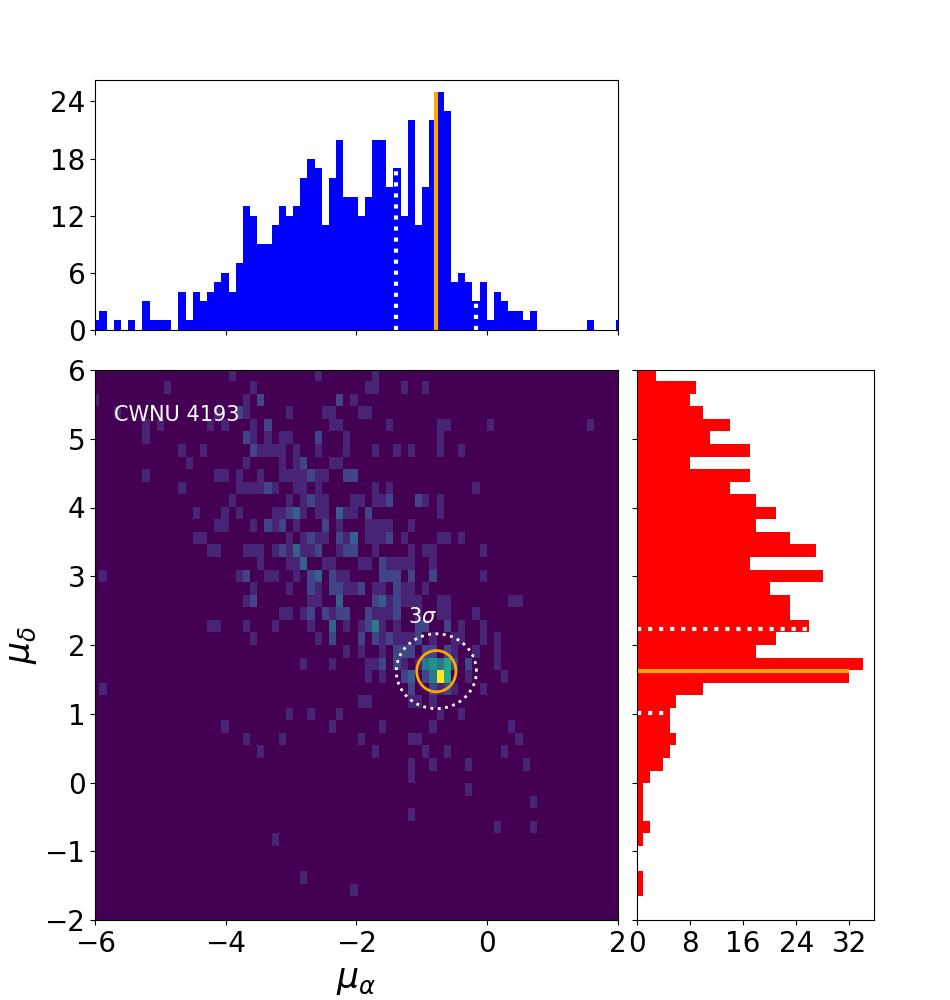}
     \end{subfigure}
    \caption{VPDs for all the stars in the respective cluster sample. VPD is the 2D histogram, along with the separate histograms for PMRA (in blue) and PMDEC (in red). The $3\sigma$ boundary centred on the mean cluster PMs are indicated by the white dotted circles. The final selection of the most likely cluster members is made within the $3~\sigma$ region, as depicted by the red circle.   
    }
    \label{ImgVPM}
\end{figure*} 
\section{Datasets} \label{datasets}
We utilize the NIR data acquired with the VISTA InfraRed CAMera (VIRCAM), from the VVVX survey. This survey was conducted at the 4.1~m wide-field Visible and Infrared Survey Telescope for Astronomy (VISTA; \citealp{2010EVISTA}) located at the European Southern Observatory (ESO) Paranal with a field of view of $1.5 deg^2$. The VVVX Survey uses the NIR passbands such as, J~(1.25 $\mu m$), H~(1.64 $\mu m$), and $K_s$~(2.14 $\mu m$) (see, \citealp{2010MinnitiVVV,VVVXDM2018,Saito+2024}). 
 
Both the VVV and VVVX photometric datasets are segmented into bulge + disk tiles. Tiles d001 to d152 in the VVV survey are for the disk region and b201 to b396 are for the bulge regions. In the VVVX regions, tiles are labelled as b401 to b512 for the bulge and e601 to e988 for the disk areas. Additionally, the VVVX area includes tiles e1001 to e1180 extending from $230^\circ to 295^\circ $ in longitude and $ -2^\circ to +2^\circ$ in latitude. The data reduction and archival merging for the VVVX Survey is  carried out by the Cambridge Astronomical Survey Unit (CASU) and by the VISTA Science Archive (VSA, \citealp{2012Cross})  at the Wide-Field Astronomy Unit (WFAU), using the VISTA Data Flow System (\citealp{2004irwin}).

Due to the high crowding of the studied regions, point-spread-function (PSF) techniques are better designed to extract the photometry of the sources there (\citealp{2018Alonso}). A new catalogue of sources for the VVVX  area in the three NIR $JHK_S$ filters using these techniques is being produced within our collaboration (Alonso-García et al., in prep), following similar steps as those described in \cite{2018Alonso} for creating the catalogue for the VVV original footprint. In this work, we use the PSF photometry from this new catalogue on the regions of the four sampled stellar clusters.

The 2MASS survey (\citealp{2MASSS2006}) is an all sky survey in the near-infrared bands J(1.25 $\mu m$), H(1.65 $\mu m$), and Ks(2.16 $\mu m$). Executed with two dedicated telescopes positioned in both hemispheres working simultaneously, achieving magnitude limits of 15.8, 15.1, and 14.3 at the J, H, and Ks bands, respectively.
To enhance our dataset, we incorporate information from the 2MASS catalogue, specifically targeting brighter stars with Ks < 11mag. This is crucial as such stars are saturated in the VVVX images. Furthermore, to ensure consistency, we convert the 2MASS photometry to the VISTA magnitude scale, as outlined by \cite{Gonz2018}, accounting for the different magnitude scales between the two photometric systems.

Gaia Data Release 3 (GDR3) includes the apparent brightness in G magnitude for more than 1.8 billion sources brighter than 21 mag, and full astrometric solution for about 1.5 billion sources. Precise PMs is provided with an accuracy of 0.02 mas/yr for sources brighter than G = 15 mag, 0.07 mas/yr for sources brighter than G = 17 mag, and 0.5 mas/yr for sources brighter than G = 20 mag (\citealp{2020Gaiacollab}). The \textit{Gaia} catalogues were subjected to quality cuts, we selected sources with astrometric excess noise $<=1.3$ mas and re-normalised unit weight error (RUWE) value of $< 1.2$. In addition we chose sources with parallax $< 0.5 mas$ to avoid severe contamination from the foreground stars.
The combination of the VVVX and 2MASS NIR data, along with the precise GAIA DR3 astrometry and PMs, enables new avenues of exploration for studying challenging environments, such as the highly reddened Milky Way bulge and disk.

The spatial density maps are constructed from the VVVX data in the Galactic coordinates as shown in Fig. \ref{ImgDmap}. We identified the apparent overdensities from the spatial density maps in these VVVX tiles e0748, e0618, e0634, and e1024. The overdensities corresponds to the star clusters FSR~1700 (\citealp{2007FSR}), FSR~1415 (\citealp{2007FSR}), Teutsch~67 (T67, \citealp{2006Kronberger}) and CWNU~4193 (C4193, \citealp{2023HeZ}). The spatial density plots for these clusters are depicted in Fig. \ref{fig:sub1}, \ref{fig:sub2}, \ref{fig:sub3} and \ref{fig:sub4}, respectively. Their central coordinates and derived Proper Motions (PMs) using the VVVX data are listed in Table \ref{tableradec}. First three clusters are poorly classified and is analysed for the first time. FSR~1415 is classified as an old open cluster by \citealp{2008Momany}. Studying these clusters in the near-infrared with the VVVX survey offers a clearer view through interstellar dust and enables precise determination of their astrophysical parameters.

\begin{table*}
    \centering
    \caption{Derived mean positions and  proper motions for the clusters}
    \label{tableradec}

    \begin{tabular}{p{1.5cm}p{2.5cm}p{3cm}p{2.5cm}p{2cm}p{2.5cm}p{1.5cm}p{2cm}}
         \hline
         \noalign{\smallskip}
         Cluster ID & RA & DEC & $\mu_\alpha$  & $\sigma_{\mu_\alpha}$ & $\mu_\delta$ & $\sigma_{\mu_\delta}$  \\
                      & [hh:mm:ss] & [dd:mm:ss] & [$mas~yr^{-1}$]& [$mas~yr^{-1}$]& [$mas~yr^{-1}$]& [$mas~yr^{-1}$]\\ 
        \noalign{\smallskip}
         \hline\hline
        \noalign{\smallskip}
         
         FSR1700       & $15:38':52.5"$      & $-59^\circ:16':03"$       & $-4.850 \pm  0.014$       & $0.250$   & $-4.030 \pm 0.013$ & $0.227$  \\ \noalign{\smallskip}
         FSR1415       & $08:40':23.9"$      & $-44^\circ:43':27"$       & $-1.971\pm0.020$       & $0.270$   & $2.305\pm0.020$ & $0.280$ \\ \noalign{\smallskip}
         Teutsch67     & $09:33':46.0"$      & $-57^\circ:05':59"$       & $-2.380\pm0.026$       & $0.223$   & $1.922\pm0.029$ & $0.245$  \\ \noalign{\smallskip}
         CWNU4193      & $08:04':41.7"$      & $-38^\circ:55':16"$       & $-0.792\pm0.025$       & $0.222$   & $1.628\pm0.213$ & $0.185$  \\ \noalign{\smallskip}

         \hline
    \end{tabular}
\end{table*}
\section{Decontamination Procedure} \label{sec3}

Exploring the inner regions of the Milky Way comes with its share of challenges. One significant factor is the differential reddening and extinction, impacting not only the accuracy of photometric distance estimates but also influencing derived cluster age and metallicity, particularly in isochrone fittings. Additionally, the presence of foreground and background field contamination significantly impacts the accuracy of our results.

To refine our analysis of the clusters, we employ a decontamination procedure that relies on PM selection, to remove the field star contamination. 
The PMs and parallaxes for the clusters are obtained from \textit{Gaia} DR3 after applying the quality cuts mentioned in the previous section. 
We initially merge the GDR3 and VVVX datasets with a matching radius of 0.5" to proceed with the decontamination procedures. We chose stars with parallaxes < 0.5 mas in order to remove the foreground population. 
After the initial selection of the cluster based on the overdensities in the spatial distribution, we analysed the Vector Point Diagram (VPD) to obtain clean samples of the clusters. The VPDs were constructed as 2D histograms and are shown in Fig. \ref{ImgVPM}.
The cluster stars are expected to show similar motions, and they are different from those of the field stars. This method will substantially reduces the field star contamination \citep[e.g.][]{2020Garro,2022Garro}.
By visually identifying the cluster peak in the VPDs using histograms as shown in Fig. \ref{ImgVPM}, and its spatial distribution we selected the cluster members that exhibits a spatial overdensity along with a peak in PMs different from that of the field. All the studied clusters show one clear overdensity in their VPDs. FSR~1415 shows a spread in overdensity in the PM space (see top-right plot in Fig. \ref{ImgVPM}) which are fluctuations in the field, and are not clustered spatially as we can see in the spatial density map of the cluster (top-right plot in Fig. \ref{ImgVPM}). 

We estimate the mean cluster PMs using the histograms of PMs in RA ($\mu_\alpha$, blue histograms) and in Dec ($\mu_\delta$, red histograms). The bin size of the histograms are the same as in the VPDs. The peak values and standard deviations ($\sigma$) of PMs for each cluster are derived by a Gaussian fitting procedure using Gaussian form in TOPCAT (\citealp{2005Topcat}). The white dotted circles in Fig. \ref{ImgVPM} represents the $3~\sigma$ regions from the cluster mean PM. The final visual selection of the most likely cluster members are taken within the $3~\sigma$ region around the peak and is indicated in the figure as the orange circle. The selected peak and $3~\sigma$ region are marked in the histograms of $\mu_\alpha$ and $\mu_\delta$ as orange solid line and white dotted lines, respectively.
Thus derived PMs and standard deviations in PMs are summarised in Table \ref{tableradec} for each candidate. 
\section{Photometric Characterisation of Clusters}

\subsection{Astrophysical Parameters}
The final catalogue obtained after decontamination is used to build the CMDs of the clusters.
The astrophysical parameters such as the age, metallicity ([Fe/H]), distance (D), reddening, and extinctions are derived from these clusters CMDs as detailed in Sect. \ref{4.1.1} and \ref{4.1.2}. The use of multiband photometry from 2MASS and Gaia along with VVVX helps us to reach more robust results. The final CMD is made by combining all the three catalogues accounting for all the quality cuts mentioned in section \ref{datasets}. The 2MASS stars in the CMD (open circles in Fig. \ref{ImgCMD}) are those bright stars ($K_s > 11 mag$) that are within the tidal radius and not detected with VVVX. The challenge was to find all the parameters simultaneously fitting both optical and NIR CMDs. 
The structural parameters of the clusters such as core radii ($r_c$), tidal radii ($r_t$), and concentration parameter ($C = log(r_t/r_c$) are also derived as detailed in Sect. \ref{4.1.3}, and are provided in Table \ref{tabfit}.

\subsubsection{Reddening, Extinction, and Distance} \label{4.1.1}

\begin{figure*}[t]
    \centering
    \begin{subfigure}[b]{0.24\textwidth}
         \includegraphics[width=\textwidth]{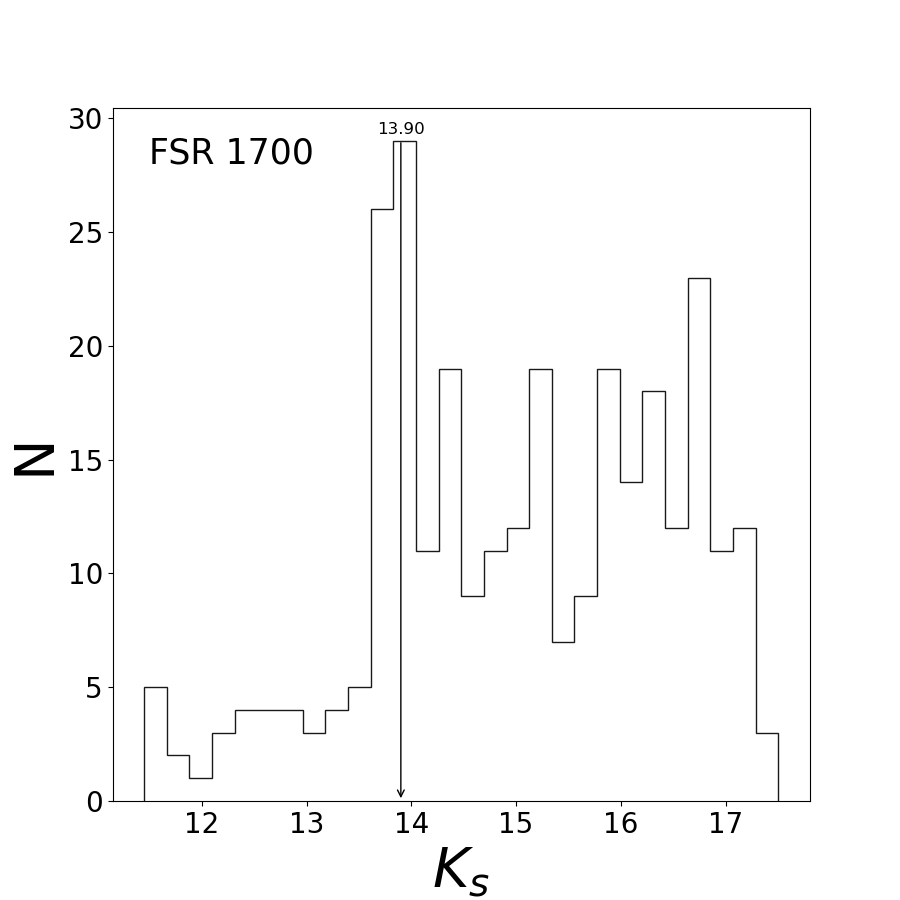}
     \end{subfigure}
      \begin{subfigure}[b]{0.24\textwidth}
         \includegraphics[width=\textwidth]{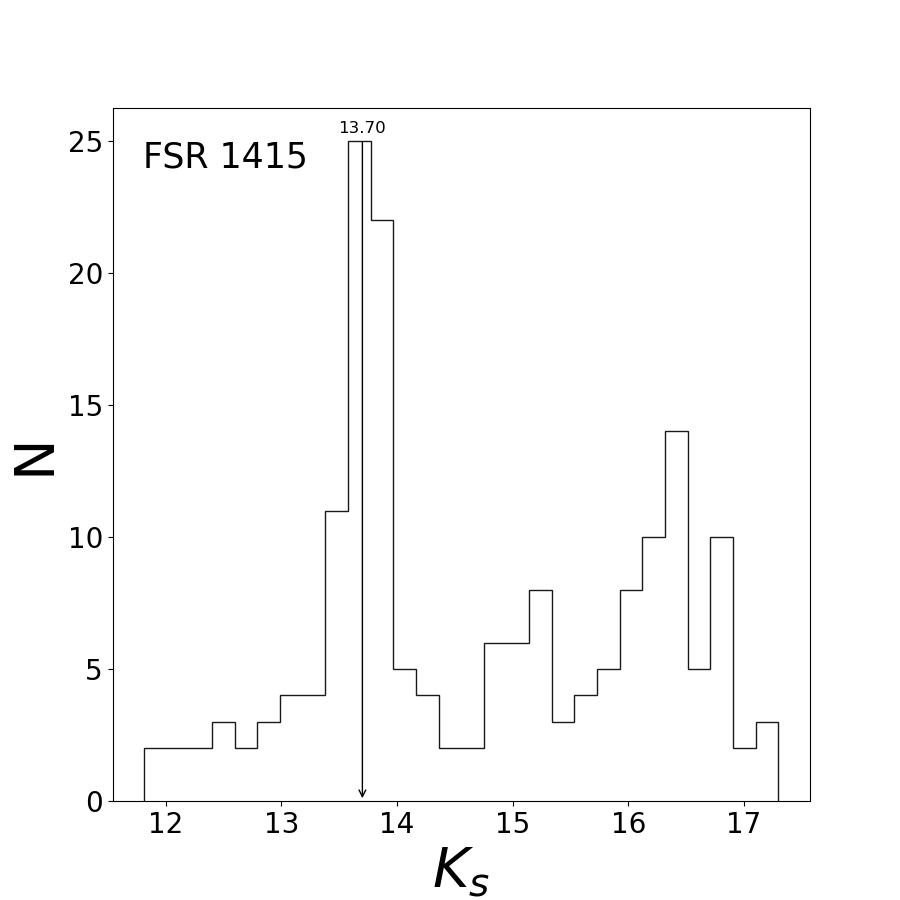}
     \end{subfigure}
     \begin{subfigure}[b]{0.24\textwidth}
         \includegraphics[width=\textwidth]{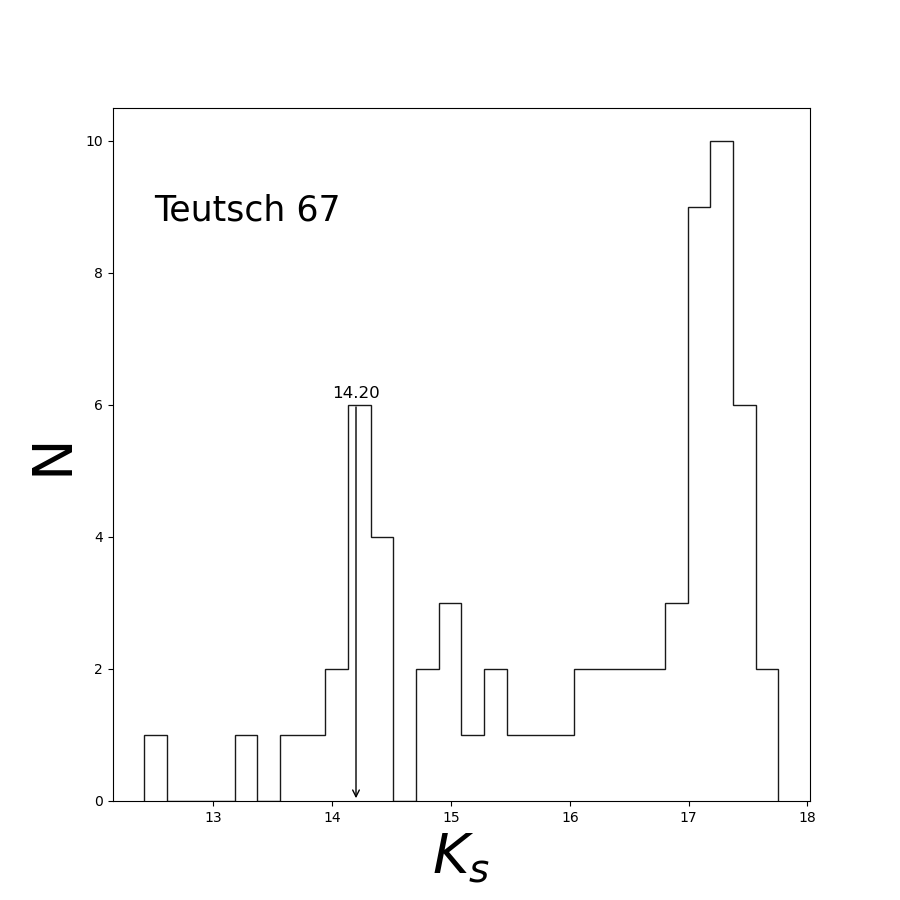}
     \end{subfigure}
      \begin{subfigure}[b]{0.24\textwidth}
         \includegraphics[width=\textwidth]{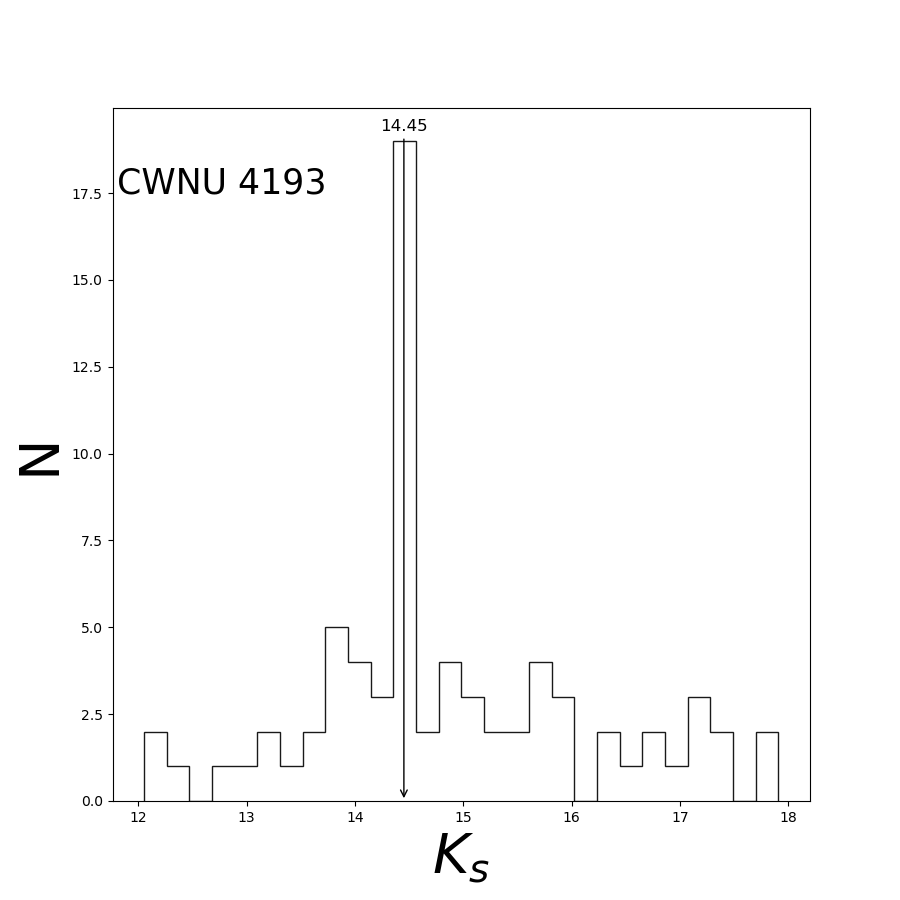}
     \end{subfigure}
    \caption{ Luminosity function in Ks-band for each cluster, showing the peaked concentrations due to the RC stars that allow us to determine accurate distances. The name of the corresponding cluster is also labelled. }
    \label{ImgLF}
\end{figure*}

In our estimation of reddening and extinction towards the clusters, we follow the approach outlined by \cite{2018Ruizdern}. Initially, we construct the luminosity function in Ks band for each cluster, as illustrated in Fig. \ref{ImgLF}. The Red Clump (RC) positions are visually identified, and are marked in the corresponding plots in Fig. \ref{ImgLF}. We then utilise the absolute magnitude of RC stars in the Ks band as $M_{Ks} = -1.605 \pm 0.009$ mag, and their intrinsic color as $(J - Ks)_0 =0.66 \pm 0.02$ mag (\citealp{2018Ruizdern}). 
The derived extinctions lies in the range $0.2 < A_{Ks} < 0.52$, and reddening in the range $0.3 < E(J-Ks) < 0.7$. 
With the aforementioned values, we calculated the heliocentric distances of the clusters by adopting the distance modulus formula. The distances vary from 9.4 kpc to 14 kpc in range.  

Additionally, we derived the extinctions and reddening in optical bands using the reddening map by \cite{2011SF}, for the mean coordinates of the RC stars. 
We then adopted the following relations for extinctions and reddening in NIR $A_{Ks} = (0.078 \pm 0.004) \times AV$ (\citealp{2019WC}). Furthermore the extinctions and reddening in the Gaia band are derived using the relations $A_G = (10.116 \pm 0.006) \times A_{ks}$, and $A_G = (1.89 \pm 0.015)\times E(BP - RP)$ (\citealp{2019WC}). We used reddening laws in optical and NIR from \cite{2019WC} to maintain the homogeneity of the analysis.

\subsubsection{Ages \& Metallicities} \label{4.1.2}

\begin{figure*}[t]
    \centering
    \begin{subfigure}[b]{0.45\textwidth}
         \includegraphics[width=\textwidth]{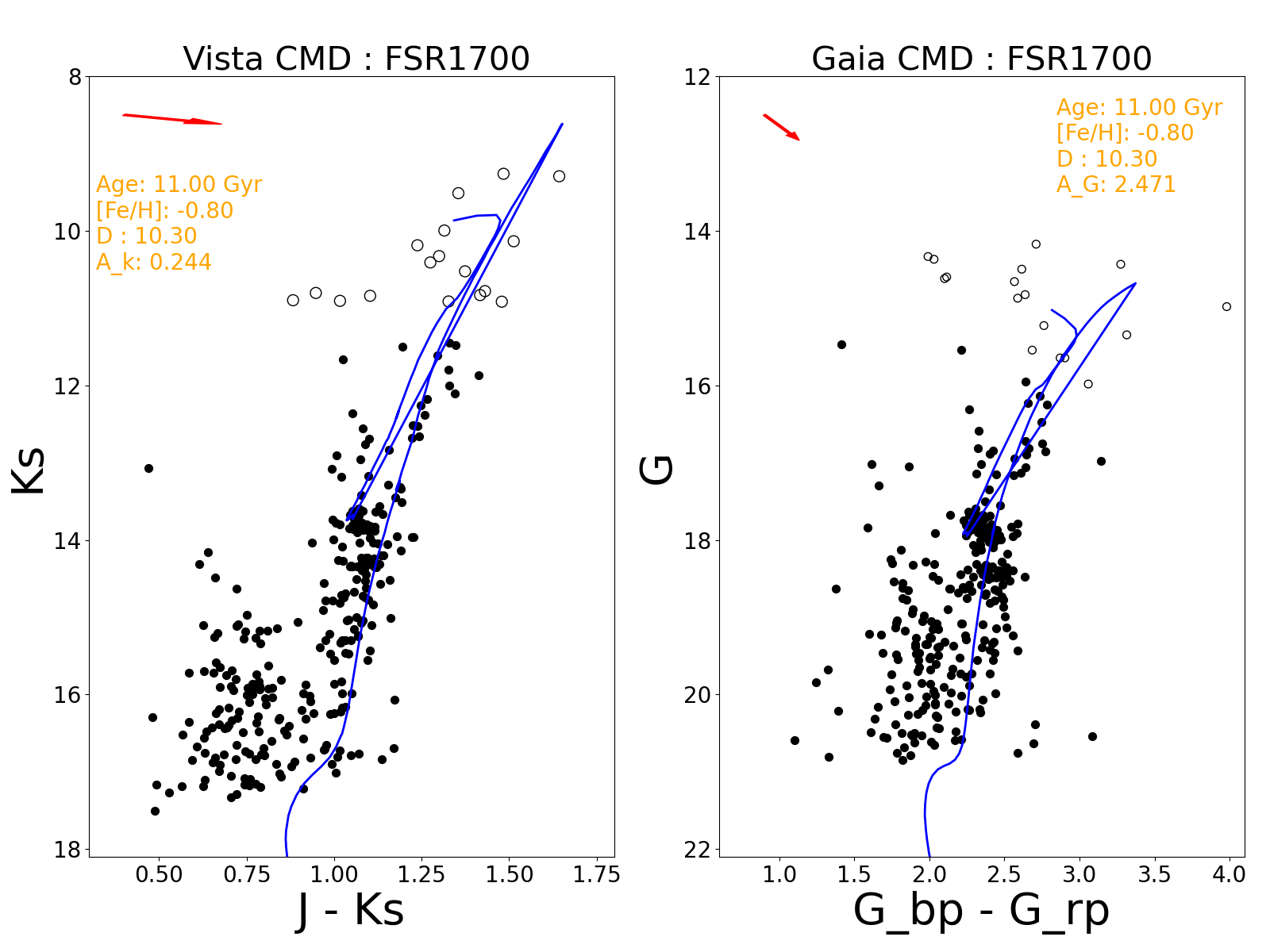}
     \end{subfigure}
      \begin{subfigure}[b]{0.45\textwidth}
         \includegraphics[width=\textwidth]{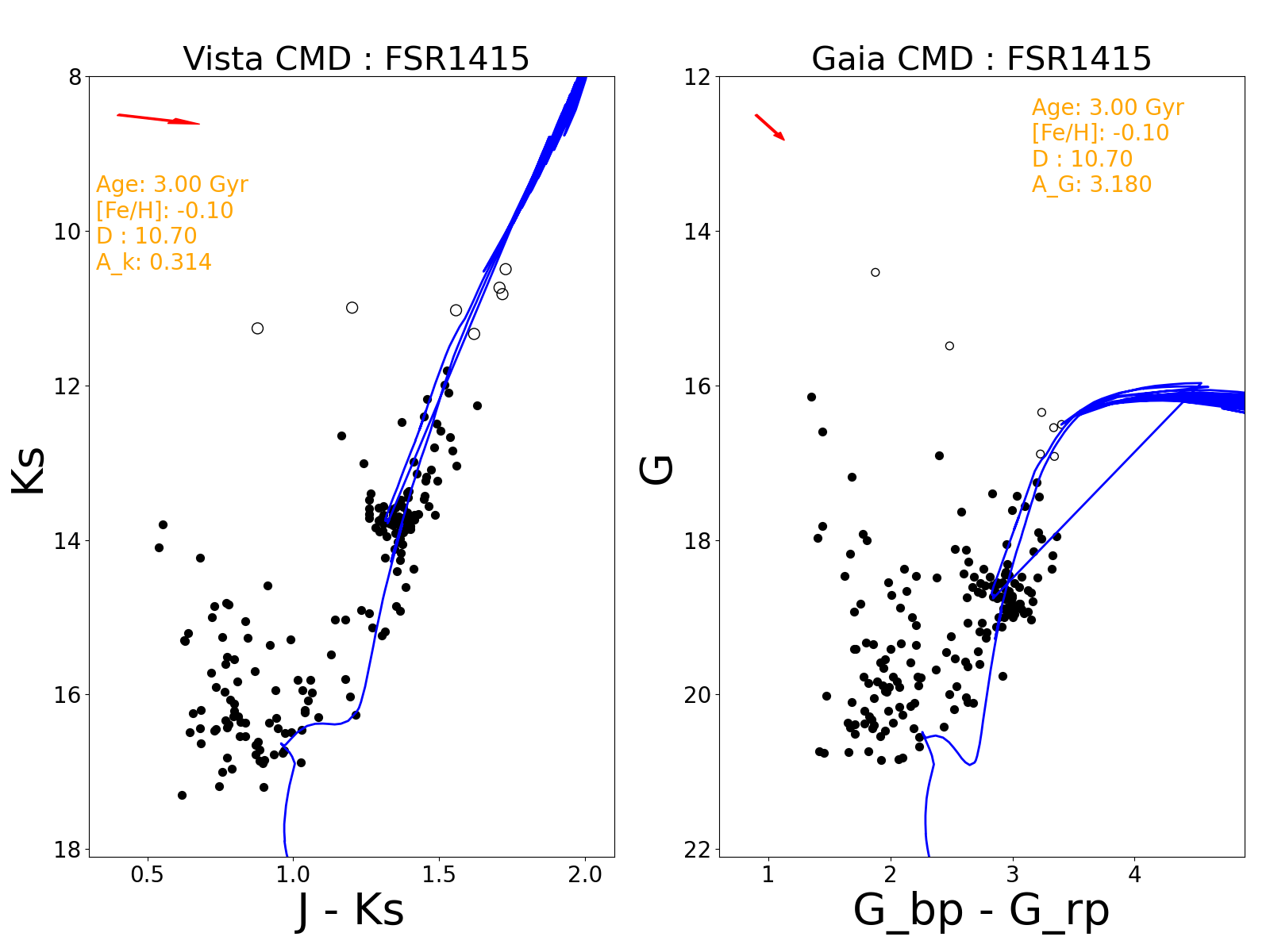}
     \end{subfigure}
     \begin{subfigure}[b]{0.45\textwidth}
         \includegraphics[width=\textwidth]{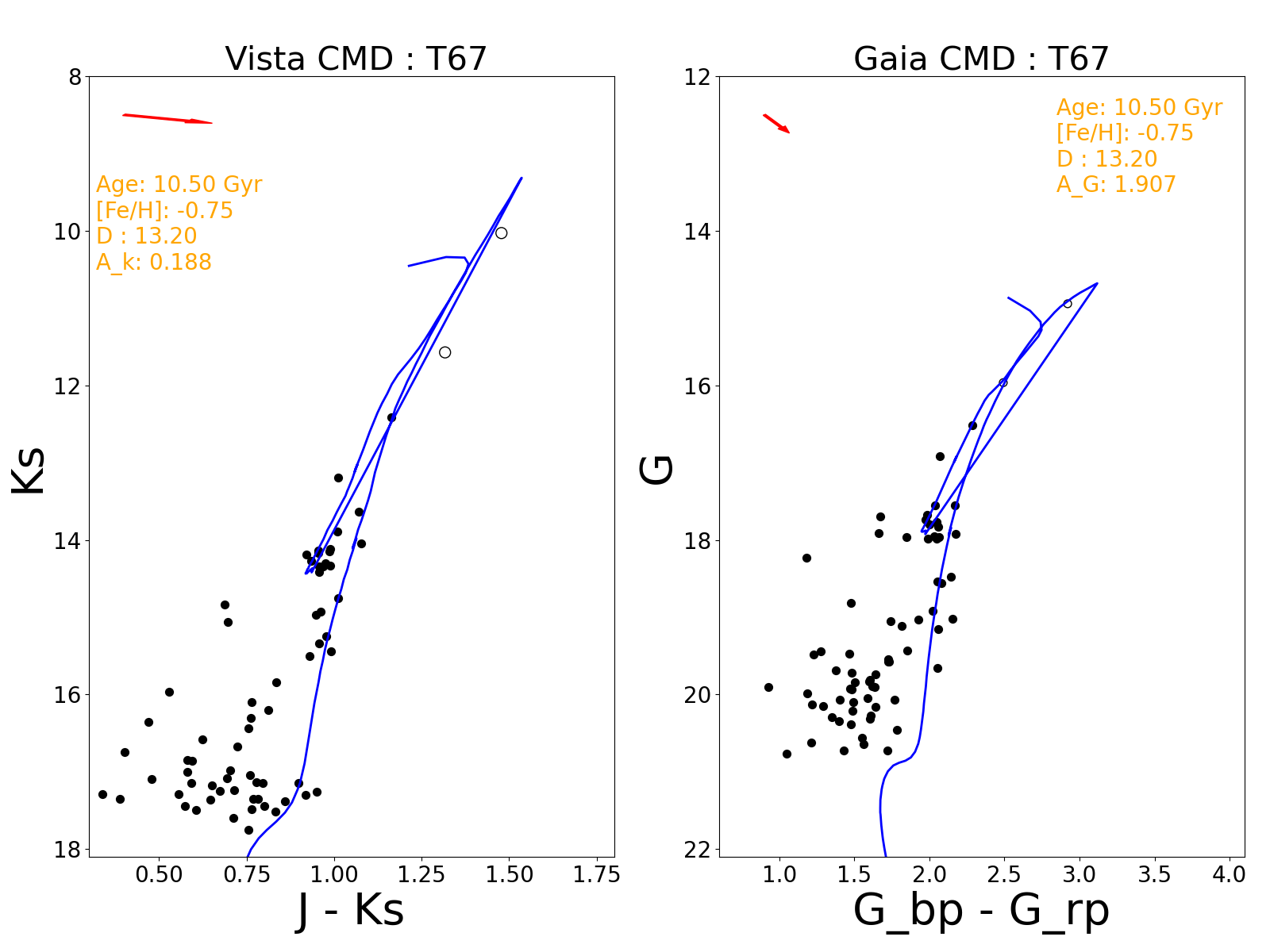}
     \end{subfigure}
      \begin{subfigure}[b]{0.45\textwidth}
         \includegraphics[width=\textwidth]{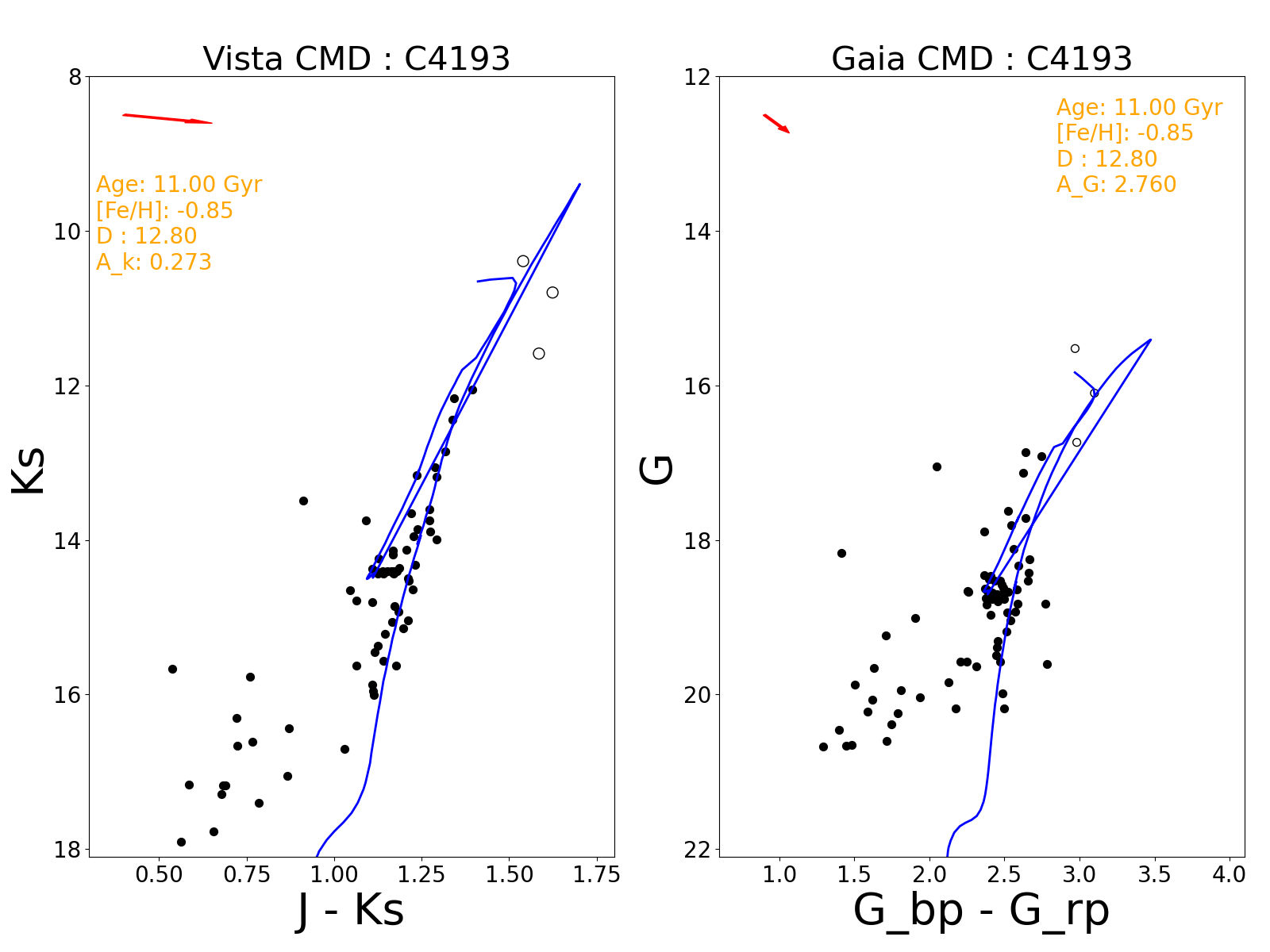}
     \end{subfigure}
    \caption{ CMDs of the clusters. For each cluster CMD on the left we have the combination of VISTA and 2MASS, whereas on the right is the Gaia CMD. The filled circles are the decontaminated cluster members from the VVVX data. The open circles in the VISTA and Gaia CMDs are the cluster members identified in the 2MASS data for the respective catalogues. The best isochrone match to the CMDs are represented by the blue solid line. The red arrows represents the reddening vectors in their respective CMDs. The errorbars in the photometry have been omitted, but the typical photometric errors are $\delta K_s = 0.01$ mag at $K_s=12$ mag,
$\delta K_s = 0.05$ mag at $K_s=16$ mag, and $\delta K_s = 0.3$ mag at $Ks=17.5$ mag,}
    \label{ImgCMD}
\end{figure*}

Considering the derived reddening, extinction and distance values, PARSEC isochrones (\citealp{Bressen2012}) are visually fitted to determine the ages and metallicities of the clusters. We fitted the isochrones simultaneously in the NIR and Optical CMDs to find the best fitted values of all astrophysical parameters simultaneously. Initially, we fixed the reddening and distance values to those estimated from the RC stars, and fitted isochrones for different ages and metallicities. Finally to obtain the best fitted isochrone we have simultaneously varied all the four parameters such as age, metallicity, distance and reddening following the procedure described in \citealp{Sar23}). Concurrently fitting the isochrone in the Gaia and VISTA CMD helps us to better constrain the slope of the RGB. Knowing the positions of the RC stars from the previous section helps to add more constrains to the visual fitting procedure. The best fitted isochrones in the VISTA and Gaia CMD are illustrated in Fig. \ref{ImgCMD}. The filled circles correspond to the decontaminated cluster members from the VVVX data and the open circles represent the brighter stars obtained from 2MASS catalogue which are above the VVVX saturation limit. The red arrow represents the derived reddening vector from the isochrone fit with an angle of $tan^{-1}(A_{K_s}/E(J-K_s)$ in the NIR CMD and $tan^{-1}(A_{G}/E(BP-RP)$ in the Gaia CMD. The best fitted parameters obtained from the isochrones are listed in Table \ref{tabfit}. The uncertainties in metallicities and distances in our study are estimated to be approximately 0.5 in ages and 0.2~dex in metallicities, consistent with findings from similar studies on GCs (e.g., \citealp{2022Garro}).

\subsection{Structural parameters} \label{4.1.3}
\begin{figure*}[t]
    \centering
    \sidecaption
    \begin{subfigure}[b]{0.45\textwidth}
         \includegraphics[width=\columnwidth]{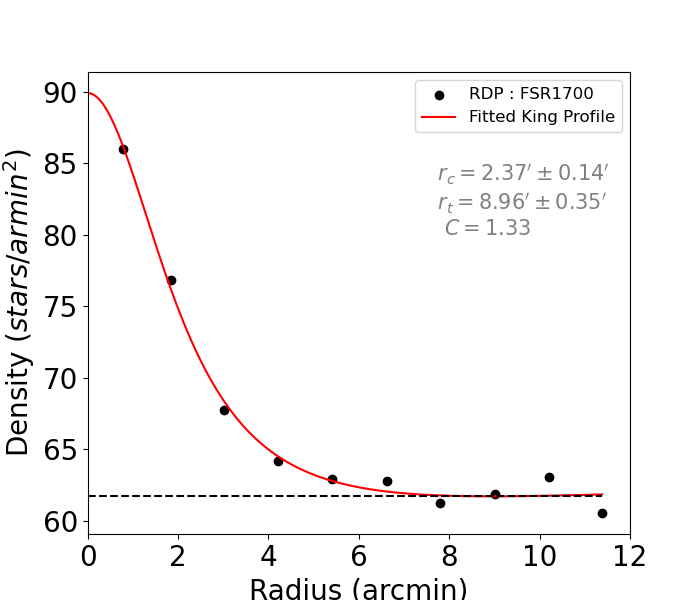}
     \end{subfigure}
      \begin{subfigure}[b]{0.45\textwidth}
         \includegraphics[width=\columnwidth]{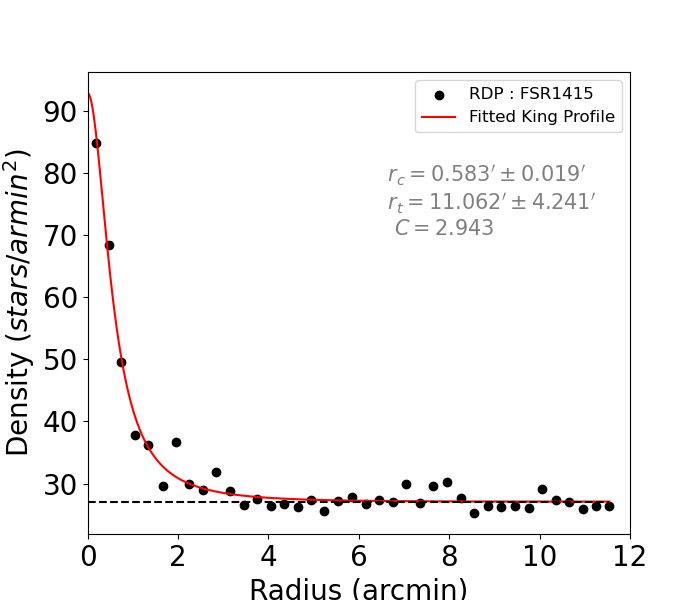}
     \end{subfigure}
     \begin{subfigure}[b]{0.465\textwidth}
         \includegraphics[width=\columnwidth]{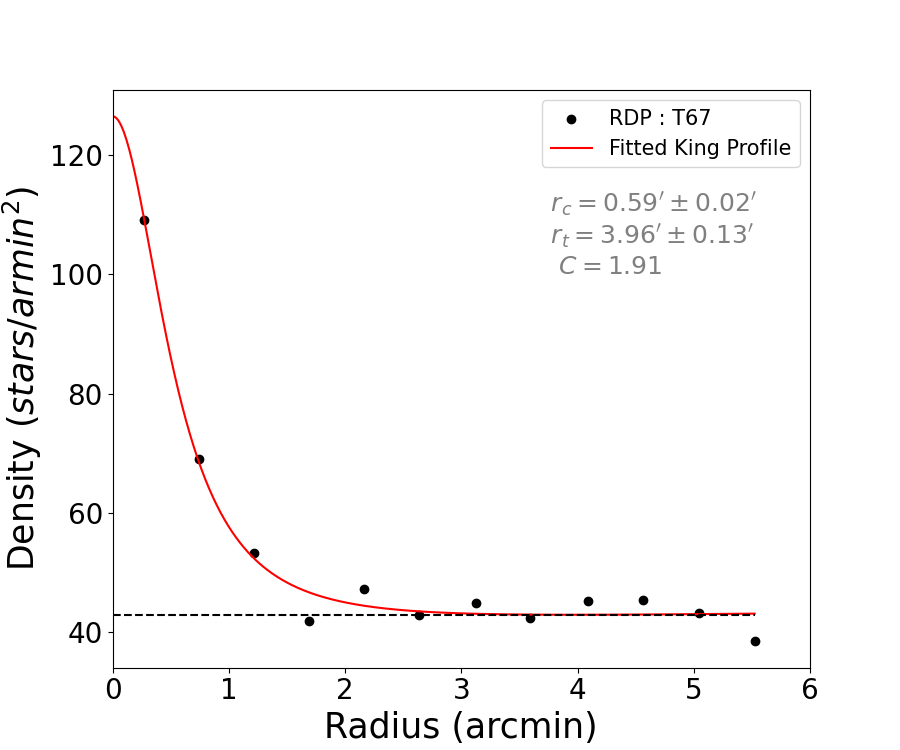}
     \end{subfigure}
      \begin{subfigure}[b]{0.45\textwidth}
         \includegraphics[width=\columnwidth]{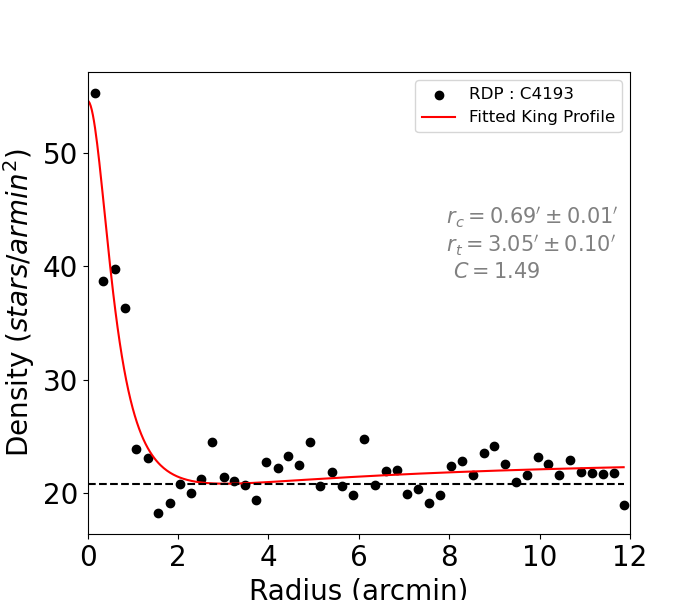}
     \end{subfigure}
    \caption{ Radial density profile of the clusters. The black points correspond to the cluster density profile, which corresponds to number of stars per unit area within an annular region. The red line shows the best-fit King model profile (\citealp{king1962structure}). The constant background density is marked with the dashed line.}
    \label{ImgRDP}
\end{figure*}

To determine the physical size of the clusters, we computed the Radial Density Profile (RDP). 
The center of the cluster is computed through a systematic procedure, beginning with initial center coordinates obtained from the literature. Subsequently, we determine the new center by calculating the median values of right ascension (RA) and declination (DEC) within a radius of 0.1' from the initial center coordinates. To ensure accuracy, we assess variations in the center coordinates by exploring different radial bins from the center.
Finally we use the center derived from the aforementioned method to construct the RDPs. The first step was dividing the sample clusters into circular different annuli with increasing radii. The number density per bin was calculated as number of stars (N) in the bin divided by the respective area (A). The RDPs for the clusters are plotted as a function of the mean distance of the circular annulus to the cluster centre to the number density in the corresponding annuli.
In our study, we employed the widely used \cite{king1962structure} model to fit the cluster density profile. We used a chisquare method to derive the best fitted model and the standard errors.
The best-fit King model gives the $r_c$, $r_t$, and $C$ of the clusters (see, Fig.\ref{ImgRDP}). 
Core collapse clusters typically exhibit higher concentration parameters compared to non-core collapse clusters (\citealp{1984Cohnandhut,Cohn+1988CCCs}). This implies that the central regions of core collapse clusters are denser and more tightly packed with stars compared to non core collapse clusters. By analysing the concentration parameters of these clusters and comparing them with known globular clusters (GCs) in the the 2010 version of the \cite{1996Harris} catalogue, it appears unlikely that FSR1700, Teutsch67, and CWNU4193 are core collapse clusters.
However the cluster FSR1415 (C $\sim 3$) could belong to the category of core collapse clusters.

\begin{table*}
    \centering
    \caption{Derived physical parameters for the observed stellar clusters.}
    \label{tabfit}

    \begin{tabular}{p{1.5cm}p{1cm}p{1.2cm}p{1.2cm}p{1cm}p{1.2cm}p{1cm}p{1.5cm} p{1.5cm}p{1.5cm}p{0.9cm}}
         \hline
         \noalign{\smallskip}
         \centering
         Cluster ID & Age & [Fe/H] & Distance  & $A_{K_s}$ & E(J-K$_s$) & $A_G$ & E(BP-RP) & $r_c$ & $r_t$ & $C$ \\
         & [Gyr] &\textbf{[dex]}&[kpc]& [mag]& [mag]& [mag]& [mag] & [arcmin]& [arcmin]& --\\ 
        \noalign{\smallskip}
         \hline\hline
        \noalign{\smallskip}
         
         FSR1700       & $\sim 11.0$      & $-0.80$       & $10.3$       & $0.24$   & $0.56$ & $2.50$ & $1.30$ & $2.37 \pm 0.14$ & $8.96 \pm 0.35$ & $1.33$ \\ \noalign{\smallskip}
         FSR1415       & $\sim 3.0$       & $-0.10$       & $10.7$       & $0.31$   & $0.73$ & $3.18$ & $1.67$ & $0.58 \pm 0.02$ & $11.06 \pm 4.2$ & $2.94$ \\ \noalign{\smallskip}
         Teutsch67     & $\sim 10.5$      & $-0.75$       & $13.2$       & $0.19$   & $0.44$ & $1.91$ & $1.01$ & $0.59 \pm 0.02$ & $3.96 \pm 0.13$ & $1.91$ \\ \noalign{\smallskip}
         CWNU4193      & $\sim 11.0$      & $-0.85$       & $12.8$       & $0.27$   & $0.63$ & $2.76$ & $1.45$ & $0.69 \pm 0.01$ & $3.05 \pm 0.10$ & $1.49$ \\ \noalign{\smallskip}

         \hline
    \end{tabular}
\end{table*}

\section{Results}
Below, we discuss the estimated parameters from the isochrone fitting for each cluster accompanied by a comparison with existing literature findings, if available. 

\subsection{FSR~1415}
The star cluster FSR~1415 was discovered by 
\cite{2007FSR}, 
located at the J2000 equatorial coordinates $\alpha = 08:40:24 $, $\delta = -44^\circ:43:05$ and galactic coordinates $l = 263.74^\circ$, $b = -1.81^\circ$. 
This cluster is identified as a genuine old open cluster by \cite{2008Momany}, using high resolution and deep photometry from the near infrared Multi-Conjugate Adaptive Optics demonstrator at the Very Large Telescope (VLT/MAD). \cite{2008Momany} derived the age, metallicity and distances using isochrones from Padova library (\citealp{2002Girardi}). Fixing a solar metallicity ($[Fe/H] = 0.0~dex$) they derived an $age = 2.5 \pm 0.7~Gyr $ and heliocentric distance $d_\odot = 8.59~kpc $, and $A_V = 4.57 $. The structural parameters of the cluster is derived using a 3-parameter King-like function, obtaining a core radius $r_c = 0.9' \pm 0.2'$, $r_t = 12.2' \pm 2.9'$ ( Fig. \ref{ImgRDP}). 
The aforementioned values derived by \cite{2008Momany} are in good agreement with our estimates of age, metallicity and distance, and the structural parameters as well. Our best fitted PARSEC isochrone yields an $age = 3 Gyr$, $[Fe/H] = 0.10 \pm 0.2~dex $, and $d = 10.7 \pm 0.5~Kpc$, $A_{Ks} = 0.31 (A_V = 4.03)$ (Fig. \ref{ImgCMD}). The concentration parameter of the cluster ( $C \sim 3$) is similar to what we find in post core-collapse GCs (\citealp{199Trager}). Moreover presence of post-core collapse features in RDPs of open clusters has been previously detected, for instance, in the $\sim1$ Gyr old cluster NGC~3960 (\citealp{2006BicaBonatto}).  

\subsection{FSR~1700}
The star cluster FSR1700 was discovered by \cite{2007FSR}, and classified as a distant reddened cluster by \cite{Buckner2013PropertiesOS}. This object was also
recently
identified as a candidate globular cluster (GC) (\citealp{2023HeZ}), and is positioned at the J2000 equatorial coordinates $\alpha = 15:38:52.5$, $\delta = -59^\circ:16:03$, and galactic coordinates $l = 322.9^\circ$, $b = -3.05^\circ$.
We utilised the Gaia PMs to derive the mean PMs of the cluster, as detailed in Section \ref{sec3}. The resulting values are $\mu_\alpha = -4.85 \pm 0.25$~$mas~yr^{-1}$ and $\mu_\delta = -4.03 \pm 0.23$~$mas~yr^{-1}$, which align well with previous findings by \cite{2023HeZ}).
The isochrone that best fits the CMD suggests an age of $11$ Gyr, a metallicity of $[Fe/H] = -0.80\pm0.2~dex$, a distance of $d = 10.3\pm0.5$~kpc, and an extinction of $A_{Ks} = 0.24$ (\ref{ImgCMD}).
The RDP was constructed using circular bins of $1$ arcmin up to a radius of $12$ arcmins. The fitted King profile yielded a core radius of $r_c = 2.37' \pm 0.14'$ and a tidal radius of $r_t = 8.96' \pm 0.35'$ (Fig .\ref{ImgRDP}). The concentration parameter, $C = 1.33$, suggests that FSR~1700 is unlikely to be a core collapse cluster (\citealp{1996Harris}).

\subsection{Teutsch~67}
The star cluster Teutsch67 was discovered by \cite{2006Kronberger},
located at the J2000 equatorial coordinates $\alpha = 09:33:46$, $\delta = -57^\circ:05:59$, and galactic coordinates $l = 278.52^\circ$, $b = -3.92^\circ$.
Utilising Gaia PMs, we determined the mean PMs of the cluster, yielding values of $\mu_\alpha = -2.38 \pm 0.03$~$mas~yr^{-1}$ and $\mu_\delta = 1.92 \pm 0.03$~$mas~yr^{-1}$.
The isochrone fitting to the CMD yields an age of $10.5$ Gyr, a metallicity of $[Fe/H] = -0.75\pm0.2~dex$, a distance of $d = 13.2\pm0.5$ kpc, and an extinction of $A_{Ks} = 0.19$ (Fig. \ref{ImgCMD}).
Constructing the RDP involved using circular bins of $0.5$ arcmin up to a radius of $6$ arcmins. The fitted King profile yielded a core radius of $r_c = 0.59' \pm 0.02'$ and a tidal radius of $r_t = 3.96' \pm 0.13'$ (Fig. \ref{ImgRDP}). Moreover, with a concentration parameter of $C = 1.91$, T67 emerges as a candidate for a GC, indicating it is unlikely to be a core collapse cluster.

\subsection{CWNU~4193}
The star cluster CWNU~4193 was recently discovered by \cite{2023HeZ},
situated at the J2000 equatorial coordinates $\alpha = 08:04:41.7$, $\delta = -38^\circ:55:16$, and galactic coordinates $l = 255.17^\circ$, $b = -3.95^\circ$.
Utilising Gaia PMs, we determined the mean PMs of the cluster, yielding values of $\mu_\alpha = -0.792 \pm 0.22$~$mas~yr^{-1}$ and $\mu_\delta = 1.63 \pm 0.21$~$mas~yr^{-1}$.
The isochrone fitting to the CMD yields an age of $11$ Gyr, a metallicity of $[Fe/H] = -0.85\pm0.2~dex$, a distance of $d = 12.8\pm0.5$ kpc, and an extinction of $A_{Ks} = 0.27$ (Fig. \ref{ImgCMD}).
Constructing the RDP involved using circular bins of $0.25$ arcmin up to a radius of $12$ arcmins. The fitted King profile yielded a core radius of $r_c = 0.69' \pm 0.01'$ and a tidal radius of $r_t = 3.05' \pm 0.10'$ (Fig. \ref{ImgRDP}). Moreover, with a concentration parameter of $C = 1.49$, CWNU~4193 emerges as a GC candidate, indicating it is unlikely to be a core collapse cluster.

\section{Summary}
In recent decades, numerous star clusters have been uncovered within the Galactic bulge; however, many remain poorly studied. In the Galactic disk, away from the bulge, open clusters are abundant while globular clusters are rare. Despite the challenges, it is crucial to discover and characterise the globular clusters existing in this region. In this study, we focus on four star clusters located in the aforementioned region, endeavouring to recover their principal astrophysical and structural parameters. To this end, we calculate the reddening and extinction across each cluster field by leveraging reddening maps in the NIR spectrum and measuring the position of the RC stars. Furthermore, we determine their distances utilising photometry from the VVV survey, supplemented by data from the Gaia mission. Employing the isochrone-fitting method with PARSEC isochrone models, we deduce their metallicities and ages.

The following three clusters, FSR~1700, Teutsch~67, and CWNU~4193, emerge as GC candidates. These clusters exhibit a concentration parameter lower than that typically observed for core collapse clusters. In contrast, FSR1415 is identified as an old open cluster showcasing post core-collapse feature, which typically displays a concentration parameter around $C \approx 3$. Our estimations of astrophysical and structural parameters for FSR~1415 align with the findings reported by \cite{2008Momany}.

We note that aside from the bulge region, the VVVX survey has been mapping the Southern Galactic plane within 230<l<350 deg, and -4.5<b<4.5 deg.
In this region of the sky covering 1080 sq.deg. there were only 3 GCs known (out of 157 objects, \citealp{1996Harris,2010Harris}). 
These previously known GCs were Lynga7 (\citealp{1995Lygna}), NGC6256, and NGC5946 (\citealp{1991NGC5946G}).
Thanks to our near-IR survey a dozen additional GCs have been identified in this region. 
These are
FSR1716=VVV-GC05 (\citealp{2017+Minniti}),
2MASS-GC 03=FSR1735 (\citealp{2016CBello}),
Garro01 (\citealp{2020Garro}),
Ferrero 54, Patchick 125, Patchick 126, Patchick122, FSR190, and Kronberger99 (\citealp{2022Garro,2023Garro}), and
FSR 1700, CWNU 2149, and Teutsch 67 (this work).
In addition, a couple more GCs were recently discovered by other surveys in this region (RLGC1 by \cite{2018Ryu}, and BH140 by \citealp{2018BH140C}). 
There are a few other objects (e.g., BH176 and ESO93-8) that are still being debated also as old open clusters. 
Therefore, the progress in the past decade has been enormous, we know now 5 times more GCs in the VVVX region of the Southern Galactic plane. 
Indeed, it is very important to have a sample of GCs as complete as possible, because old GCs can be used to trace the formation history of our Galaxy, revealing past events of accretion of dwarf galaxies (e.g., \citealp{2019Massari,2020Forbes,2021VasiBaum}). 
Based on these results, we can expect that a deeper and higher resolution near-IR survey of the whole Galactic plane made with the future Roman Space Telescope (\citealp{2023RomanTele}) would find dozens of new GCs at these low latitudes, some of which may reveal concealed accretion events.

\section*{Acknowledgements}

We gratefully acknowledge data from the ESO Public Survey programs 179.B-2002 and 198.B-2004 taken with the VISTA telescope, and products from the Cambridge Astronomical Survey Unit (CASU) and the Wide Field Astronomy Unit at the Royal Observatory, Edinburgh.
DM thanks the support from the ANID BASAL Center for Astrophysics and Associated Technologies (CATA) projects ACE210002 and FB210003, from Fondecyt Regular No. 1220724, and from CNPq Brasil Project 350104/2022-0.
This research was partially supported by the Argentinian institution SECYT (Universidad
Nacional de Córdoba) and Consejo Nacional de Investigaciones Científicas y Técnicas de la República Argentina, Agencia Nacional de
Promoción Científica y Tecnológica.
BD acknowledges support by ANID-FONDECYT iniciación grant No. 11221366 and from the ANID Basal project FB210003.
J.A.-G. acknowledges support by Fondecyt Regular 1201490 and ANID – Millennium Science Initiative Program – ICN12\_009 awarded to the Millennium Institute of
Astrophysics (MAS).

\bibliography{biblio.bib}

\begin{thebibliography}{52}
\expandafter\ifx\csname natexlab\endcsname\relax\def\natexlab#1{#1}\fi

\bibitem[{{Alcaino} {et~al.}(1991){Alcaino}, {Liller}, {Alvarado}, \& {Wenderoth}}]{1991NGC5946G}
{Alcaino}, G., {Liller}, W., {Alvarado}, F., \& {Wenderoth}, E. 1991, \aj, 102, 1371

\bibitem[{{Alonso-Garc{\'\i}a} {et~al.}(2018){Alonso-Garc{\'\i}a}, {Saito}, {Hempel}, {Minniti}, {Pullen}, {Catelan}, {Ramos}, {Cross}, {Gonzalez}, {Lucas}, {Palma}, {Valenti}, \& {Zoccali}}]{2018Alonso}
{Alonso-Garc{\'\i}a}, J., {Saito}, R.~K., {Hempel}, M., {et~al.} 2018, \aap, 619, A4

\bibitem[{{Bica} {et~al.}(2006){Bica}, {Bonatto}, \& {Blumberg}}]{2006BicaBonatto}
{Bica}, E., {Bonatto}, C., \& {Blumberg}, R. 2006, \aap, 460, 83

\bibitem[{{Bica} {et~al.}(2019){Bica}, {Pavani}, {Bonatto}, \& {Lima}}]{2019+Bica}
{Bica}, E., {Pavani}, D.~B., {Bonatto}, C.~J., \& {Lima}, E.~F. 2019, \aj, 157, 12

\bibitem[{{Bressan} {et~al.}(2012){Bressan}, {Marigo}, {Girardi}, {Salasnich}, {Dal Cero}, {Rubele}, \& {Nanni}}]{Bressen2012}
{Bressan}, A., {Marigo}, P., {Girardi}, L., {et~al.} 2012, \mnras, 427, 127

\bibitem[{Buckner \& Froebrich(2013)}]{Buckner2013PropertiesOS}
Buckner, A. S.~M. \& Froebrich, D. 2013, Monthly Notices of the Royal Astronomical Society, 436, 1465

\bibitem[{{Cantat-Gaudin} {et~al.}(2018){Cantat-Gaudin}, {Vallenari}, {Sordo}, {Pensabene}, {Krone-Martins}, {Moitinho}, {Jordi}, {Casamiquela}, {Balaguer-N{\'u}nez}, {Soubiran}, \& {Brouillet}}]{2018BH140C}
{Cantat-Gaudin}, T., {Vallenari}, A., {Sordo}, R., {et~al.} 2018, \aap, 615, A49

\bibitem[{{Carballo-Bello} {et~al.}(2016){Carballo-Bello}, {Ram{\'\i}rez Alegr{\'\i}a}, {Borissova}, {Smith}, {Kurtev}, {Lucas}, {Moni Bidin}, {Alonso-Garc{\'\i}a}, {Minniti}, {Palma}, {D{\'e}k{\'a}ny}, {Medina}, {Moyano}, {Villanueva}, \& {Kuhn}}]{2016CBello}
{Carballo-Bello}, J.~A., {Ram{\'\i}rez Alegr{\'\i}a}, S., {Borissova}, J., {et~al.} 2016, \mnras, 462, 501

\bibitem[{Cohn {et~al.}(1988)Cohn, Grindlay, Bailyn, \& Hertz}]{Cohn+1988CCCs}
Cohn, H., Grindlay, J., Bailyn, C., \& Hertz, P. 1988, 126, 657

\bibitem[{{Cross} {et~al.}(2012){Cross}, {Collins}, {Mann}, {Read}, {Sutorius}, {Blake}, {Holliman}, {Hambly}, {Emerson}, {Lawrence}, \& {Noddle}}]{2012Cross}
{Cross}, N.~J.~G., {Collins}, R.~S., {Mann}, R.~G., {et~al.} 2012, \aap, 548, A119

\bibitem[{{Emerson} \& {Sutherland}(2010)}]{2010EVISTA}
{Emerson}, J. \& {Sutherland}, W. 2010, The Messenger, 139, 2

\bibitem[{{Forbes} {et~al.}(2020){Forbes}, {Alabi}, {Romanowsky}, {Brodie}, \& {Arimoto}}]{2020Forbes}
{Forbes}, D.~A., {Alabi}, A., {Romanowsky}, A.~J., {Brodie}, J.~P., \& {Arimoto}, N. 2020, \mnras, 492, 4874

\bibitem[{{Froebrich} {et~al.}(2007){Froebrich}, {Scholz}, \& {Raftery}}]{2007FSR}
{Froebrich}, D., {Scholz}, A., \& {Raftery}, C.~L. 2007, \mnras, 374, 399

\bibitem[{{Gaia Collaboration}(2020)}]{2020Gaiacollab}
{Gaia Collaboration}. 2020, {VizieR Online Data Catalog: Gaia EDR3 (Gaia Collaboration, 2020)}, VizieR On-line Data Catalog: I/350. Originally published in: 2020A\&A...649A...1G; doi:10.5270/esa-1ug

\bibitem[{{Gaia Collaboration} {et~al.}(2023){Gaia Collaboration}, {Vallenari}, {Brown}, {Prusti}, {de Bruijne}, {Arenou}, {Babusiaux}, {Biermann}, {Creevey}, {Ducourant}, {Evans}, {Eyer}, {Guerra}, {Hutton}, {Jordi}, {Klioner}, {Lammers}, {Lindegren}, {Luri}, {Mignard}, {Panem}, {Pourbaix}, {Randich}, {Sartoretti}, {Soubiran}, {Tanga}, {Walton}, {Bailer-Jones}, {Bastian}, {Drimmel}, {Jansen}, {Katz}, {Lattanzi}, {van Leeuwen}, {Bakker}, {Cacciari}, {Casta{\~n}eda}, {De Angeli}, {Fabricius}, {Fouesneau}, {Fr{\'e}mat}, {Galluccio}, {Guerrier}, {Heiter}, {Masana}, {Messineo}, {Mowlavi}, {Nicolas}, {Nienartowicz}, {Pailler}, {Panuzzo}, {Riclet}, {Roux}, {Seabroke}, {Sordo}, {Th{\'e}venin}, {Gracia-Abril}, {Portell}, {Teyssier}, {Altmann}, {Andrae}, {Audard}, {Bellas-Velidis}, {Benson}, {Berthier}, {Blomme}, {Burgess}, {Busonero}, {Busso}, {C{\'a}novas}, {Carry}, {Cellino}, {Cheek}, {Clementini}, {Damerdji}, {Davidson}, {de Teodoro}, {Nu{\~n}ez Campos}, {Delchambre}, {Dell'Oro}, {Esquej},
  {Fern{\'a}ndez-Hern{\'a}ndez}, {Fraile}, {Garabato}, {Garc{\'\i}a-Lario}, {Gosset}, {Haigron}, {Halbwachs}, {Hambly}, {Harrison}, {Hern{\'a}ndez}, {Hestroffer}, {Hodgkin}, {Holl}, {Jan{\ss}en}, {Jevardat de Fombelle}, {Jordan}, {Krone-Martins}, {Lanzafame}, {L{\"o}ffler}, {Marchal}, {Marrese}, {Moitinho}, {Muinonen}, {Osborne}, {Pancino}, {Pauwels}, {Recio-Blanco}, {Reyl{\'e}}, {Riello}, {Rimoldini}, {Roegiers}, {Rybizki}, {Sarro}, {Siopis}, {Smith}, {Sozzetti}, {Utrilla}, {van Leeuwen}, {Abbas}, {{\'A}brah{\'a}m}, {Abreu Aramburu}, {Aerts}, {Aguado}, {Ajaj}, {Aldea-Montero}, {Altavilla}, {{\'A}lvarez}, {Alves}, {Anders}, {Anderson}, {Anglada Varela}, {Antoja}, {Baines}, {Baker}, {Balaguer-N{\'u}{\~n}ez}, {Balbinot}, {Balog}, {Barache}, {Barbato}, {Barros}, {Barstow}, {Bartolom{\'e}}, {Bassilana}, {Bauchet}, {Becciani}, {Bellazzini}, {Berihuete}, {Bernet}, {Bertone}, {Bianchi}, {Binnenfeld}, {Blanco-Cuaresma}, {Blazere}, {Boch}, {Bombrun}, {Bossini}, {Bouquillon}, {Bragaglia}, {Bramante}, {Breedt},
  {Bressan}, {Brouillet}, {Brugaletta}, {Bucciarelli}, {Burlacu}, {Butkevich}, {Buzzi}, {Caffau}, {Cancelliere}, {Cantat-Gaudin}, {Carballo}, {Carlucci}, {Carnerero}, {Carrasco}, {Casamiquela}, {Castellani}, {Castro-Ginard}, {Chaoul}, {Charlot}, {Chemin}, {Chiaramida}, {Chiavassa}, {Chornay}, {Comoretto}, {Contursi}, {Cooper}, {Cornez}, {Cowell}, {Crifo}, {Cropper}, {Crosta}, {Crowley}, {Dafonte}, {Dapergolas}, {David}, {David}, {de Laverny}, {De Luise}, {De March}, {De Ridder}, {de Souza}, {de Torres}, {del Peloso}, {del Pozo}, {Delbo}, {Delgado}, {Delisle}, {Demouchy}, {Dharmawardena}, {Di Matteo}, {Diakite}, {Diener}, {Distefano}, {Dolding}, {Edvardsson}, {Enke}, {Fabre}, {Fabrizio}, {Faigler}, {Fedorets}, {Fernique}, {Fienga}, {Figueras}, {Fournier}, {Fouron}, {Fragkoudi}, {Gai}, {Garcia-Gutierrez}, {Garcia-Reinaldos}, {Garc{\'\i}a-Torres}, {Garofalo}, {Gavel}, {Gavras}, {Gerlach}, {Geyer}, {Giacobbe}, {Gilmore}, {Girona}, {Giuffrida}, {Gomel}, {Gomez}, {Gonz{\'a}lez-N{\'u}{\~n}ez},
  {Gonz{\'a}lez-Santamar{\'\i}a}, {Gonz{\'a}lez-Vidal}, {Granvik}, {Guillout}, {Guiraud}, {Guti{\'e}rrez-S{\'a}nchez}, {Guy}, {Hatzidimitriou}, {Hauser}, {Haywood}, {Helmer}, {Helmi}, {Sarmiento}, {Hidalgo}, {Hilger}, {H{\l}adczuk}, {Hobbs}, {Holland}, {Huckle}, {Jardine}, {Jasniewicz}, {Jean-Antoine Piccolo}, {Jim{\'e}nez-Arranz}, {Jorissen}, {Juaristi Campillo}, {Julbe}, {Karbevska}, {Kervella}, {Khanna}, {Kontizas}, {Kordopatis}, {Korn}, {K{\'o}sp{\'a}l}, {Kostrzewa-Rutkowska}, {Kruszy{\'n}ska}, {Kun}, {Laizeau}, {Lambert}, {Lanza}, {Lasne}, {Le Campion}, {Lebreton}, {Lebzelter}, {Leccia}, {Leclerc}, {Lecoeur-Taibi}, {Liao}, {Licata}, {Lindstr{\o}m}, {Lister}, {Livanou}, {Lobel}, {Lorca}, {Loup}, {Madrero Pardo}, {Magdaleno Romeo}, {Managau}, {Mann}, {Manteiga}, {Marchant}, {Marconi}, {Marcos}, {Marcos Santos}, {Mar{\'\i}n Pina}, {Marinoni}, {Marocco}, {Marshall}, {Martin Polo}, {Mart{\'\i}n-Fleitas}, {Marton}, {Mary}, {Masip}, {Massari}, {Mastrobuono-Battisti}, {Mazeh}, {McMillan}, {Messina}, {Michalik},
  {Millar}, {Mints}, {Molina}, {Molinaro}, {Moln{\'a}r}, {Monari}, {Mongui{\'o}}, {Montegriffo}, {Montero}, {Mor}, {Mora}, {Morbidelli}, {Morel}, {Morris}, {Muraveva}, {Murphy}, {Musella}, {Nagy}, {Noval}, {Oca{\~n}a}, {Ogden}, {Ordenovic}, {Osinde}, {Pagani}, {Pagano}, {Palaversa}, {Palicio}, {Pallas-Quintela}, {Panahi}, {Payne-Wardenaar}, {Pe{\~n}alosa Esteller}, {Penttil{\"a}}, {Pichon}, {Piersimoni}, {Pineau}, {Plachy}, {Plum}, {Poggio}, {Pr{\v{s}}a}, {Pulone}, {Racero}, {Ragaini}, {Rainer}, {Raiteri}, {Rambaux}, {Ramos}, {Ramos-Lerate}, {Re Fiorentin}, {Regibo}, {Richards}, {Rios Diaz}, {Ripepi}, {Riva}, {Rix}, {Rixon}, {Robichon}, {Robin}, {Robin}, {Roelens}, {Rogues}, {Rohrbasser}, {Romero-G{\'o}mez}, {Rowell}, {Royer}, {Ruz Mieres}, {Rybicki}, {Sadowski}, {S{\'a}ez N{\'u}{\~n}ez}, {Sagrist{\`a} Sell{\'e}s}, {Sahlmann}, {Salguero}, {Samaras}, {Sanchez Gimenez}, {Sanna}, {Santove{\~n}a}, {Sarasso}, {Schultheis}, {Sciacca}, {Segol}, {Segovia}, {S{\'e}gransan}, {Semeux}, {Shahaf}, {Siddiqui}, {Siebert},
  {Siltala}, {Silvelo}, {Slezak}, {Slezak}, {Smart}, {Snaith}, {Solano}, {Solitro}, {Souami}, {Souchay}, {Spagna}, {Spina}, {Spoto}, {Steele}, {Steidelm{\"u}ller}, {Stephenson}, {S{\"u}veges}, {Surdej}, {Szabados}, {Szegedi-Elek}, {Taris}, {Taylor}, {Teixeira}, {Tolomei}, {Tonello}, {Torra}, {Torra}, {Torralba Elipe}, {Trabucchi}, {Tsounis}, {Turon}, {Ulla}, {Unger}, {Vaillant}, {van Dillen}, {van Reeven}, {Vanel}, {Vecchiato}, {Viala}, {Vicente}, {Voutsinas}, {Weiler}, {Wevers}, {Wyrzykowski}, {Yoldas}, {Yvard}, {Zhao}, {Zorec}, {Zucker}, \& {Zwitter}}]{GDR32023}
{Gaia Collaboration}, {Vallenari}, A., {Brown}, A.~G.~A., {et~al.} 2023, \aap, 674, A1

\bibitem[{{Garro} {et~al.}(2023){Garro}, {Fern{\'a}ndez-Trincado}, {Minniti}, {Moya}, {Palma}, {Beers}, {Placco}, {Barbuy}, {Sneden}, {Alves-Brito}, {Dias}, {Af{\c{s}}ar}, {Frelijj}, \& {Lane}}]{2023Garro}
{Garro}, E.~R., {Fern{\'a}ndez-Trincado}, J.~G., {Minniti}, D., {et~al.} 2023, \aap, 669, A136

\bibitem[{{Garro} {et~al.}(2020){Garro}, {Minniti}, {G{\'o}mez}, {Alonso-Garc{\'\i}a}, {Barb{\'a}}, {Barbuy}, {Clari{\'a}}, {Chen{\'e}}, {Dias}, {Hempel}, {Ivanov}, {Lucas}, {Majaess}, {Mauro}, {Moni Bidin}, {Palma}, {Pullen}, {Saito}, {Smith}, {Surot}, {Ram{\'\i}rez Alegr{\'\i}a}, {Rejkuba}, {Ripepi}, \& {Fern{\'a}ndez Trincado}}]{2020Garro}
{Garro}, E.~R., {Minniti}, D., {G{\'o}mez}, M., {et~al.} 2020, \aap, 642, L19

\bibitem[{{Garro} {et~al.}(2022){Garro}, {Minniti}, {G{\'o}mez}, {Alonso-Garc{\'\i}a}, {Ripepi}, {Fern{\'a}ndez-Trincado}, \& {Vivanco C{\'a}diz}}]{2022Garro}
{Garro}, E.~R., {Minniti}, D., {G{\'o}mez}, M., {et~al.} 2022, \aap, 658, A120

\bibitem[{{Girardi} {et~al.}(2002){Girardi}, {Bertelli}, {Bressan}, {Chiosi}, {Groenewegen}, {Marigo}, {Salasnich}, \& {Weiss}}]{2002Girardi}
{Girardi}, L., {Bertelli}, G., {Bressan}, A., {et~al.} 2002, \aap, 391, 195

\bibitem[{{Gonzalez} {et~al.}(2012){Gonzalez}, {Rejkuba}, {Zoccali}, {Valenti}, {Minniti}, {Schultheis}, {Tobar}, \& {Chen}}]{2012Gonzalez}
{Gonzalez}, O.~A., {Rejkuba}, M., {Zoccali}, M., {et~al.} 2012, \aap, 543, A13

\bibitem[{{Gonz{\'a}lez-Fern{\'a}ndez} {et~al.}(2018){Gonz{\'a}lez-Fern{\'a}ndez}, {Hodgkin}, {Irwin}, {Gonz{\'a}lez-Solares}, {Koposov}, {Lewis}, {Emerson}, {Hewett}, {Yolda{\c{s}}}, \& {Riello}}]{Gonz2018}
{Gonz{\'a}lez-Fern{\'a}ndez}, C., {Hodgkin}, S.~T., {Irwin}, M.~J., {et~al.} 2018, \mnras, 474, 5459

\bibitem[{{Gran} {et~al.}(2019){Gran}, {Zoccali}, {Contreras Ramos}, {Valenti}, {Rojas-Arriagada}, {Carballo-Bello}, {Alonso-Garcia}, {Minniti}, {Rejkuba}, \& {Surot}}]{2019Gran}
{Gran}, F., {Zoccali}, M., {Contreras Ramos}, R., {et~al.} 2019, \aap, 628, A45

\bibitem[{{Harris}(1996)}]{1996Harris}
{Harris}, W.~E. 1996, \aj, 112, 1487

\bibitem[{{Harris}(2010)}]{2010Harris}
{Harris}, W.~E. 2010, arXiv e-prints, arXiv:1012.3224

\bibitem[{{He} {et~al.}(2023){He}, {Luo}, {Wang}, {Ren}, {Peng}, {Cui}, {Liu}, \& {Jiang}}]{2023HeZ}
{He}, Z., {Luo}, Y., {Wang}, K., {et~al.} 2023, \apjs, 267, 34

\bibitem[{{Irwin} {et~al.}(2004){Irwin}, {Lewis}, {Hodgkin}, {Bunclark}, {Evans}, {McMahon}, {Emerson}, {Stewart}, \& {Beard}}]{2004irwin}
{Irwin}, M.~J., {Lewis}, J., {Hodgkin}, S., {et~al.} 2004, in Society of Photo-Optical Instrumentation Engineers (SPIE) Conference Series, Vol. 5493, Optimizing Scientific Return for Astronomy through Information Technologies, ed. P.~J. {Quinn} \& A.~{Bridger}, 411--422

\bibitem[{King(1962)}]{king1962structure}
King, I. 1962, Astronomical Journal, Vol. 67, p. 471 (1962), 67, 471

\bibitem[{{Kronberger} {et~al.}(2006){Kronberger}, {Teutsch}, {Alessi}, {Steine}, {Ferrero}, {Graczewski}, {Juchert}, {Patchick}, {Riddle}, {Saloranta}, {Schoenball}, \& {Watson}}]{2006Kronberger}
{Kronberger}, M., {Teutsch}, P., {Alessi}, B., {et~al.} 2006, \aap, 447, 921

\bibitem[{{Massari} {et~al.}(2019){Massari}, {Koppelman}, \& {Helmi}}]{2019Massari}
{Massari}, D., {Koppelman}, H.~H., \& {Helmi}, A. 2019, \aap, 630, L4

\bibitem[{{Minniti}(2018)}]{VVVXDM2018}
{Minniti}, D. 2018, in Astrophysics and Space Science Proceedings, Vol.~51, The Vatican Observatory, Castel Gandolfo: 80th Anniversary Celebration, ed. G.~{Gionti} \& J.-B. {Kikwaya Eluo}, 63

\bibitem[{{Minniti} {et~al.}(2021{\natexlab{a}}){Minniti}, {Fern{\'a}ndez-Trincado}, {G{\'o}mez}, {Smith}, {Lucas}, \& {Contreras Ramos}}]{2021+DM}
{Minniti}, D., {Fern{\'a}ndez-Trincado}, J.~G., {G{\'o}mez}, M., {et~al.} 2021{\natexlab{a}}, \aap, 650, L11

\bibitem[{{Minniti} {et~al.}(2017{\natexlab{a}}){Minniti}, {Geisler}, {Alonso-Garc{\'\i}a}, {Palma}, {Beam{\'\i}n}, {Borissova}, {Catelan}, {Clari{\'a}}, {Cohen}, {Contreras Ramos}, {Dias}, {Fern{\'a}ndez-Trincado}, {G{\'o}mez}, {Hempel}, {Ivanov}, {Kurtev}, {Lucas}, {Moni-Bidin}, {Pullen}, {Ram{\'\i}rez Alegr{\'\i}a}, {Saito}, \& {Valenti}}]{2017DM}
{Minniti}, D., {Geisler}, D., {Alonso-Garc{\'\i}a}, J., {et~al.} 2017{\natexlab{a}}, \apjl, 849, L24

\bibitem[{{Minniti} {et~al.}(2010){Minniti}, {Lucas}, {Emerson}, {Saito}, {Hempel}, {Pietrukowicz}, {Ahumada}, {Alonso}, {Alonso-Garcia}, {Arias}, {Bandyopadhyay}, {Barb{\'a}}, {Barbuy}, {Bedin}, {Bica}, {Borissova}, {Bronfman}, {Carraro}, {Catelan}, {Clari{\'a}}, {Cross}, {de Grijs}, {D{\'e}k{\'a}ny}, {Drew}, {Fari{\~n}a}, {Feinstein}, {Fern{\'a}ndez Laj{\'u}s}, {Gamen}, {Geisler}, {Gieren}, {Goldman}, {Gonzalez}, {Gunthardt}, {Gurovich}, {Hambly}, {Irwin}, {Ivanov}, {Jord{\'a}n}, {Kerins}, {Kinemuchi}, {Kurtev}, {L{\'o}pez-Corredoira}, {Maccarone}, {Masetti}, {Merlo}, {Messineo}, {Mirabel}, {Monaco}, {Morelli}, {Padilla}, {Palma}, {Parisi}, {Pignata}, {Rejkuba}, {Roman-Lopes}, {Sale}, {Schreiber}, {Schr{\"o}der}, {Smith}, {}, {Soto}, {Tamura}, {Tappert}, {Thompson}, {Toledo}, {Zoccali}, \& {Pietrzynski}}]{2010MinnitiVVV}
{Minniti}, D., {Lucas}, P.~W., {Emerson}, J.~P., {et~al.} 2010, \na, 15, 433

\bibitem[{{Minniti} {et~al.}(2021{\natexlab{b}}){Minniti}, {Palma}, \& {Clari{\'a}}}]{2021DMinniti}
{Minniti}, D., {Palma}, T., \& {Clari{\'a}}, J.~J. 2021{\natexlab{b}}, Boletin de la Asociacion Argentina de Astronomia La Plata Argentina, 62, 107

\bibitem[{{Minniti} {et~al.}(2017{\natexlab{b}}){Minniti}, {Palma}, {D{\'e}k{\'a}ny}, {Hempel}, {Rejkuba}, {Pullen}, {Alonso-Garc{\'\i}a}, {Barb{\'a}}, {Barbuy}, {Bica}, {Bonatto}, {Borissova}, {Catelan}, {Carballo-Bello}, {Chene}, {Clari{\'a}}, {Cohen}, {Contreras Ramos}, {Dias}, {Emerson}, {Froebrich}, {Buckner}, {Geisler}, {Gonzalez}, {Gran}, {Hajdu}, {Irwin}, {Ivanov}, {Kurtev}, {Lucas}, {Majaess}, {Mauro}, {Moni-Bidin}, {Navarrete}, {Ram{\'\i}rez Alegr{\'\i}a}, {Saito}, {Valenti}, \& {Zoccali}}]{2017+Minniti}
{Minniti}, D., {Palma}, T., {D{\'e}k{\'a}ny}, I., {et~al.} 2017{\natexlab{b}}, \apjl, 838, L14

\bibitem[{{Momany} {et~al.}(2008){Momany}, {Ortolani}, {Bonatto}, {Bica}, \& {Barbuy}}]{2008Momany}
{Momany}, Y., {Ortolani}, S., {Bonatto}, C., {Bica}, E., \& {Barbuy}, B. 2008, \mnras, 391, 1650

\bibitem[{{Obasi} {et~al.}(2021){Obasi}, {G{\'o}mez}, {Minniti}, \& {Alonso-Garc{\'\i}a}}]{2021casmir}
{Obasi}, C., {G{\'o}mez}, M., {Minniti}, D., \& {Alonso-Garc{\'\i}a}, J. 2021, \aap, 654, A39

\bibitem[{{Paladini} {et~al.}(2023){Paladini}, {Zucker}, {Benjamin}, {Nataf}, {Minniti}, {Zasowski}, {Peek}, {Carey}, {Allen}, {Alonso-Garcia}, {Alves}, {Anders}, {Athanassoula}, {Beers}, {Bird}, {Bland-Hwathorn}, {Brown}, {Buder}, {Casagrande}, {Casey}, {Cassisi}, {Catelan}, {Chary}, {Chene}, {Ciardi}, {Comeron}, {Cohen}, {Dame}, {Drimmel}, {Fernandez Trincado}, {Finkbeiner}, {Geisler}, {Gennaro}, {Goodman}, {Green}, {Hajdu}, {Henderson}, {Hora}, {Ivanov}, {Kirkpatrick}, {Kobayashi}, {Kuhn}, {Kunder}, {Lu}, {Lucas}, {Majaess}, {Megeath}, {Meisner}, {Molinari}, {Mroz}, {Ness}, {Neumayer}, {Nogueras-Lara}, {Noriega-Crespo}, {Poleski}, {Rix}, {Rebull}, {Reggiani}, {Rejkuba}, {Saito}, {Schoenrich}, {Saydjari}, {Schisano}, {Schlafly}, {Schlaufman}, {Smith}, {Speagle}, {Wisz}, {Wyse}, \& {Zakamska}}]{2023RomanTele}
{Paladini}, R., {Zucker}, C., {Benjamin}, R., {et~al.} 2023, arXiv e-prints, arXiv:2307.07642

\bibitem[{{Palma} {et~al.}(2019){Palma}, {Minniti}, {Alonso-Garc{\'\i}a}, {Crestani}, {Netzel}, {Clari{\'a}}, {Saito}, {Dias}, {Fern{\'a}ndez-Trincado}, {Kammers}, {Geisler}, {G{\'o}mez}, {Hempel}, \& {Pullen}}]{2019Palma}
{Palma}, T., {Minniti}, D., {Alonso-Garc{\'\i}a}, J., {et~al.} 2019, \mnras, 487, 3140

\bibitem[{{Ruiz-Dern} {et~al.}(2018){Ruiz-Dern}, {Babusiaux}, {Arenou}, {Turon}, \& {Lallement}}]{2018Ruizdern}
{Ruiz-Dern}, L., {Babusiaux}, C., {Arenou}, F., {Turon}, C., \& {Lallement}, R. 2018, \aap, 609, A116

\bibitem[{{Ryu} \& {Lee}(2018)}]{2018Ryu}
{Ryu}, J. \& {Lee}, M.~G. 2018, \apjl, 863, L38

\bibitem[{Saito {et~al.}(2024)Saito, Hempel, \& Alonso-García}]{Saito+2024}
Saito, R.~K., Hempel, M., \& Alonso-García, J. 2024, The VISTA Variables in the Vía Láctea eXtended ESO public survey (VVVX): completion of the observations and legacy., submitted to A\&A

\bibitem[{{Saito} {et~al.}(2012){Saito}, {Hempel}, {Minniti}, {Lucas}, {Rejkuba}, {Toledo}, {Gonzalez}, {Alonso-Garc{\'\i}a}, {Irwin}, {Gonzalez-Solares}, {Hodgkin}, {Lewis}, {Cross}, {Ivanov}, {Kerins}, {Emerson}, {Soto}, {Am{\^o}res}, {Gurovich}, {D{\'e}k{\'a}ny}, {Angeloni}, {Beamin}, {Catelan}, {Padilla}, {Zoccali}, {Pietrukowicz}, {Moni Bidin}, {Mauro}, {Geisler}, {Folkes}, {Sale}, {Borissova}, {Kurtev}, {Ahumada}, {Alonso}, {Adamson}, {Arias}, {Bandyopadhyay}, {Barb{\'a}}, {Barbuy}, {Baume}, {Bedin}, {Bellini}, {Benjamin}, {Bica}, {Bonatto}, {Bronfman}, {Carraro}, {Chen{\`e}}, {Clari{\'a}}, {Clarke}, {Contreras}, {Corvill{\'o}n}, {de Grijs}, {Dias}, {Drew}, {Fari{\~n}a}, {Feinstein}, {Fern{\'a}ndez-Laj{\'u}s}, {Gamen}, {Gieren}, {Goldman}, {Gonz{\'a}lez-Fern{\'a}ndez}, {Grand}, {Gunthardt}, {Hambly}, {Hanson}, {He{\l}miniak}, {Hoare}, {Huckvale}, {Jord{\'a}n}, {Kinemuchi}, {Longmore}, {L{\'o}pez-Corredoira}, {Maccarone}, {Majaess}, {Mart{\'\i}n}, {Masetti}, {Mennickent}, {Mirabel}, {Monaco}, {Morelli},
  {Motta}, {Palma}, {Parisi}, {Parker}, {Pe{\~n}aloza}, {Pietrzy{\'n}ski}, {Pignata}, {Popescu}, {Read}, {Rojas}, {Roman-Lopes}, {Ruiz}, {Saviane}, {Schreiber}, {Schr{\"o}der}, {Sharma}, {Smith}, {Sodr{\'e}}, {Stead}, {Stephens}, {Tamura}, {Tappert}, {Thompson}, {Valenti}, {Vanzi}, {Walton}, {Weidmann}, \& {Zijlstra}}]{2012Saito}
{Saito}, R.~K., {Hempel}, M., {Minniti}, D., {et~al.} 2012, \aap, 537, A107

\bibitem[{{Saroon} {et~al.}(2023){Saroon}, {Dias}, {Tsujimoto}, {Parisi}, {Maia}, {Kerber}, {Bekki}, {Minniti}, {Oliveira}, {Westera}, {Katime Santrich}, {Bica}, {Sanmartim}, {Correa Quint}, \& {Fraga}}]{Sar23}
{Saroon}, S., {Dias}, B., {Tsujimoto}, T., {et~al.} 2023, \aap, 677, A35

\bibitem[{{Schlafly} \& {Finkbeiner}(2011)}]{2011SF}
{Schlafly}, E.~F. \& {Finkbeiner}, D.~P. 2011, \apj, 737, 103

\bibitem[{{Skrutskie} {et~al.}(2006){Skrutskie}, {Cutri}, {Stiening}, {Weinberg}, {Schneider}, {Carpenter}, {Beichman}, {Capps}, {Chester}, {Elias}, {Huchra}, {Liebert}, {Lonsdale}, {Monet}, {Price}, {Seitzer}, {Jarrett}, {Kirkpatrick}, {Gizis}, {Howard}, {Evans}, {Fowler}, {Fullmer}, {Hurt}, {Light}, {Kopan}, {Marsh}, {McCallon}, {Tam}, {Van Dyk}, \& {Wheelock}}]{2MASSS2006}
{Skrutskie}, M.~F., {Cutri}, R.~M., {Stiening}, R., {et~al.} 2006, \aj, 131, 1163

\bibitem[{Smith {et~al.}(2018)Smith, Lucas, Kurtev, Smart, Minniti, Borissova, Jones, Zhang, Marocco, Contreras~Pena, {et~al.}}]{smith2018virac}
Smith, L., Lucas, P., Kurtev, R., {et~al.} 2018, Monthly Notices of the Royal Astronomical Society, 474, 1826

\bibitem[{{Tavarez} \& {Friel}(1995)}]{1995Lygna}
{Tavarez}, M. \& {Friel}, E.~D. 1995, \aj, 110, 223

\bibitem[{{Taylor}(2005)}]{2005Topcat}
{Taylor}, M.~B. 2005, in Astronomical Society of the Pacific Conference Series, Vol. 347, Astronomical Data Analysis Software and Systems XIV, ed. P.~{Shopbell}, M.~{Britton}, \& R.~{Ebert}, 29

\bibitem[{{Trager} {et~al.}(1995){Trager}, {King}, \& {Djorgovski}}]{199Trager}
{Trager}, S.~C., {King}, I.~R., \& {Djorgovski}, S. 1995, \aj, 109, 218

\bibitem[{{Vasiliev} \& {Baumgardt}(2021)}]{2021VasiBaum}
{Vasiliev}, E. \& {Baumgardt}, H. 2021, \mnras, 505, 5978

\bibitem[{{Wang} \& {Chen}(2019)}]{2019WC}
{Wang}, S. \& {Chen}, X. 2019, \apj, 877, 116

\end{thebibliography}

\end{document}